\documentclass[reprint]{revtex4-2}

\usepackage{dcolumn}
\usepackage{bm}

\usepackage{amssymb}
\usepackage{amsmath}
\usepackage{mathtools}
\usepackage{bm}
\usepackage{subfig}
\usepackage{xcolor}
\usepackage{makecell}
\usepackage{float}
\usepackage[export]{adjustbox}
\usepackage{mwe}
\usepackage{booktabs}
\usepackage{graphicx}

\begin{document}

\preprint{APS/123-QED}

\title{Machine Learning Enhanced Collision Operator \\for the Lattice Boltzmann Method \\Based on Invariant Networks}

\author{Mario Christopher Bedrunka}
\email{bedrunka.research@gmail.com}
\affiliation{Chair of Fluid Mechanics, University of Siegen, Paul-Bonatz-Straße 9-11, 57076 Siegen-Weidenau, Germany}
\affiliation{Institute of Technology, Resource and Energy-efficient Engineering (TREE),\\ Bonn-Rhein-Sieg University of Applied Sciences,
Grantham-Allee 20, 53757 Sankt Augustin, Germany}

\author{Tobias Horstmann}
\affiliation{Department of Engine Acoustics, German Aerospace Center, Bismarckstraße 101, 10625 Berlin, Germany}

\author{Ben Picard}
\affiliation{Institute of Technology, Resource and Energy-efficient Engineering (TREE),\\ Bonn-Rhein-Sieg University of Applied Sciences,
Grantham-Allee 20, 53757 Sankt Augustin, Germany}

\author{Dirk Reith}
\affiliation{Institute of Technology, Resource and Energy-efficient Engineering (TREE),\\ Bonn-Rhein-Sieg University of Applied Sciences,
Grantham-Allee 20, 53757 Sankt Augustin, Germany}
\affiliation{Fraunhofer Institute for Algorithms and Scientific Computing (SCAI), Schloss Birlinghoven, 53754 Sankt Augustin, Germany}

\author{Holger Foysi}
\affiliation{Chair of Fluid Mechanics, University of Siegen, Paul-Bonatz-Straße 9-11, 57076 Siegen-Weidenau, Germany}

\date{\today}

\begin{abstract}
Integrating machine learning techniques in established numerical solvers represents a modern approach to enhancing computational fluid dynamics simulations. Within the lattice Boltzmann method  (\textsc{lbm}), the collision operator serves as an ideal entry point to incorporate machine learning techniques to enhance its accuracy and stability. In this work, an invariant neural network is constructed, acting on an equivariant collision operator, optimizing the relaxation rates of non-physical moments. This optimization deliberately enhances robustness to symmetry transformations and ensures consistent behavior across geometric operations. The proposed neural collision operator (\textsc{nco}) is trained using forced isotropic turbulence simulations driven by spectral forcing, ensuring stable turbulence statistics. The desired performance is achieved by minimizing the energy spectrum discrepancy between direct numerical simulations and underresolved simulations over a specified wave number range. The loss function is further extended to tailor numerical dissipation at high wave numbers, ensuring robustness without compromising accuracy at low and intermediate wave numbers. The \textsc{nco}'s performance is demonstrated using three-dimensional Taylor-Green vortex (\textsc{tgv}) flows, where it accurately predicts the dynamics even in highly underresolved simulations. Compared to other \textsc{lbm} models, such as the \textsc{bgk} and \textsc{kbc} operators, the \textsc{nco} exhibits superior accuracy while maintaining stability. In addition, the operator shows robust performance in alternative configurations, including turbulent three-dimensional cylinder flow. Finally, an alternative training procedure using time-dependent quantities is introduced. It is based on a reduced \textsc{tgv} model along with newly proposed symmetry boundary conditions. The reduction in memory consumption enables training at significantly higher Reynolds numbers, successfully leading to stable yet accurate simulations.
\end{abstract}


\maketitle

\section{Introduction}\label{sec:introduction}
Turbulent flows are ubiquitous and play a key role in various physical phenomena, ranging from atmospheric dynamics and ocean currents to aeroacoustics and industrial applications \cite{Kov:1953,Canuto1998,DOE_COMB,pope2000,WangM:2006, falcucci2021}. Hence, understanding and accurately predicting the behavior of turbulent flows is paramount for numerous scientific and engineering fields. However, because of their inherently complex nature, turbulent flows pose significant challenges. Direct numerical simulations (\textsc{dns}) require immense computational resources and are therefore not suitable for most realistic flows of interest in engineering \cite{Coleman2010}. Traditionally, scientists bypass complexity by making use of Reynolds-averaged Navier-Stokes modeling (\textsc{rans}) or large-eddy simulations (\textsc{les}), among others.
An alternative way to solve the problem is to utilize a discrete version of the Boltzmann equation on a regular grid with a simplified collision operator, referred to as Lattice Boltzmann method {(\textsc{lbm})} \cite{Succi1991}. Contrary to the methods mentioned above, \textsc{lbm} simulates fluid behavior through collision and streaming of particle distribution functions on this discrete grid. Despite advantages regarding parallelization, locality, and simplicity, many of the aforementioned issues remain.

Recent studies have shown promising potential in enhancing numerical methods for simulating turbulent flows by integrating neural networks \cite{brunton2020}. These advanced computational tools learn from data, enabling them to predict complex flow dynamics more efficiently than traditional methods. For example, neural networks (\textsc{nn}) can refine simulations by correcting errors in coarse-grid computations \cite{wang2017}, offer improved models for unresolved scales in large eddy simulations \cite{maulik2019}, and even directly construct turbulence dynamics \cite{kim2019}. Using machine learning to identify complex patterns in the data, these approaches can potentially overcome some limitations of traditional methods. Combining machine learning with classical computational fluid dynamics (\textsc{cfd}) can thus provide alternative methods for analyzing turbulent phenomena that are both more efficient and potentially more accurate than traditional approaches.

In recent decades, the (\textsc{lbm}), as mentioned above, has emerged as a powerful alternative computational tool for studying fluid dynamics \cite{chen1998, kruger2017}. The fundamental algorithm of the lattice Boltzmann equation has been further enriched by recent studies. These include, among many others, off-lattice Boltzmann methods to separate the coupling between time and space discretization \cite{Kraemer:2017,Kramer2020,Wilde2021}. Other developments were designed to tackle multiscale flows, such as the discrete unified gas kinetic scheme (\textsc{dugks}) \cite{Guo2021}, or to capture interfaces in multiphase flows by combining a Lax-Wendroff propagation scheme with a specific equilibrium distribution function \cite{Lou2015}. Several \textsc{lbm} methods were developed to provide stability even at higher Reynolds numbers or in under-resolved situations \cite{Ricot2009,Karlin2014,Geier2015,Kraemer2019,Hosseini2023}. Most recently, first steps were taken to explore the integration of neural networks in \textsc{lbm} methods \cite{bedrunka2021,corbetta2023, horstmann2024,ortali2024,gabbana2024}. These neural networks can enhance flow simulations by adding correction values to the numerical method to enhance accuracy in simulations of under-resolved flow (e.g., cylinder flow \cite{ataei2024}) or by predicting optimal relaxation parameters to ensure predetermined behavior \cite{bedrunka2021}. This is contrary to classical approaches often used for optimization (e.g. \cite{Marinc2012}). The main goal here is to achieve an optimal balance between stability and accuracy, addressing the limitations of the standard \textsc{lbm} method.

A collision operator decisively constitutes the solver's physics of the \textsc{lbm}. It determines the number of degrees of freedom represented by hydrodynamic moments and higher-order moments. The latter is usually propagated in such a way as to enable them to improve accuracy and stability.  These models are well known as multiple relaxation time operators \cite{Karlin2014, lallemand2000, dellar2003}. To examine the relaxation rates required for these models, Simons et al. \cite{simonis2021} identified optimal sets by brute-force analysis to conclude stability criteria in turbulent flows. However, these sets are chosen as constant values through the simulation and do not adapt to dynamic changes including laminar-turbulent transition. In this work, we want to take advantage of propagating higher-order moments to define a neural collision operator \textsc{nco} as an adaptive model to calculate the relaxation rates depending on the local flow state.

Recently, a PyTorch-based \textsc{lbm} framework was developed to easily incorporate \textsc{ml} algorithms into \textsc{lbm} \cite{bedrunka2021}. The present paper makes use of this framework to combine \textsc{ml} with \textsc{lbm} to develop a novel neural collision operator \textsc{nco}, allowing accurate and robust three-dimensional simulations of turbulence. Additionally, leveraging the inherent symmetries of the stencils, an invariant neural network is constructed to provide predictions regardless of orientation.

This paper is structured as follows: Section \ref{sec:methodology} describes the methodology, which comprises the lattice Boltzmann method in its fundamental form, the invariant neural network, and the coupling by introducing a \textsc{nco} with equivariant properties. Section \ref{sec:results} shows the training results and the performance of the \textsc{nco}. Then, benchmarks are presented that evaluate the accuracy of various flow simulations (i.e., forced isotropic turbulence, Taylor-Green vortex, and three-dimensional cylinder flow). Section \ref{sec:conclusion} constitutes the conclusion.

\section{Methodology} \label{sec:methodology}
\subsection{Lattice Boltzmann Method} \label{sec:lbm}
The lattice Boltzmann method (\textsc{lbm}) is derived from the Boltzmann equation, which is discretized to describe a particle distribution function $\boldsymbol f = (f_0,f_1,...f_i,...f_{q-1})$ as
\begin{equation}
    \begin{split}
    f_{i}\left(\mathbf{x}+\mathbf{c}_{i}\delta_t, t+\delta_t \right) = f_{i}\left(\mathbf{x},t \right) + \Omega_{i}\left(f_{i}(\mathbf{x}, t) \right) + S_{i}\left(f_{i}(\mathbf{x}, t) \right), \\
    i \in \{0,1,\dots q-1\}.
    \end{split}
\end{equation}
The symbols $\mathbf{c}_i$ and $\delta_t$ define the discrete particle velocities and the time step, respectively \cite{kruger2017}. The fundamental method based on two operations. First, a collision operator $\boldsymbol\Omega:\mathbb{R}^q\to\mathbb{R}^q$ models the particle interaction on each node, which leads to a local redistribution of the particle distribution function, where $q$ is the number of discrete particle velocities, determined by the stencil $\text{D}d\text{Q}q$, including the zero velocity. The symbol $d$ describes the number of dimensions. Second, the streaming step moves the distributions along their trajectories. The most frequently used collision operator is the Bhatnagar-Gross-Krook (\textsc{bgk}) model, which is written for the components as
\begin{equation}
    \Omega_{i}\left(\boldsymbol{f}\right) = -\frac{f_{i}-f_{i}^{\text{eq}}}{\tau}.
\end{equation}
This operator relaxes the distribution function towards its discrete quadratic equilibrium, as defined by
\begin{equation} \label{Eq:EquilibriumDistribution}
    f_{i}^{\text{eq}}\left(\mathbf{x},t\right) = w_{i}\rho\left( 1 + \frac{\mathbf{u \cdot c}_{i}}{c_{s}^{2}} + \frac{\left(\mathbf{u \cdot c}_{i}\right)^{2}}{2c_{s}^{4}} - \frac{\mathbf{u}\cdot\mathbf{u}}{2c_{s}^{2}} \right),
\end{equation}
where $w_{i}$ and $c_s$ are the lattice weights and the speed of sound, respectively. The approach to equilibrium is modulated by a single relaxation parameter $\tau$, which controls the rate of relaxation. The density $\rho$ and the fluid velocity $\mathbf{u}$ are obtained as
\begin{equation}
    \rho \left( \mathbf{x}, t\right) = \sum_{i} f_{i} \left(\mathbf{x},t \right)
    \qquad \mathrm{and} \qquad
    \rho \mathbf{u}\left( \mathbf{x}, t\right) = \sum_{i} \mathbf{c}_{i}f_{i} \left(\mathbf{x},t \right) .
    \vspace{-.3cm}
\end{equation}

Incorporating forces into the lattice Boltzmann method can be achieved using the source term $S_{i}$ in the lattice Boltzmann equation. External effects can be represented, enabling the modeling of energy injection into the system and the generation of accelerating forces that impact the flow dynamics. This study utilizes the proposed force method of Kupershtokh \cite{kupershtokh2009} who proposed a scheme that shifts $f_{i}$ in velocity space such that:
\begin{equation}
   S_{i} = f_{i}^{\text{eq}}\left( \rho, \mathbf{u}+\Delta \mathbf{u}\right) - f_{i}^{\text{eq}}\left( \rho, \mathbf{u}\right),
\end{equation}
where $\Delta \mathbf{u}$ describes the velocity shift within the equilibrium distribution and is obtained using the force $\mathbf{F}$ as
\begin{equation}
    \Delta \mathbf{u} = \mathbf{F} \Delta t / \rho.
\end{equation}

\subsection{Neural Collision Operator} \label{sec:network}
\begin{figure*}[t]
    \centering
    \includegraphics[width=1\linewidth]{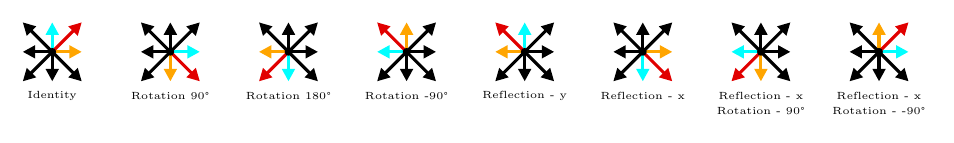}
    \vspace{-1cm}
    \caption{Illustration of group operations applied on the $D2Q9$ stencil, revealing new orientations.}
    \label{fig:symmetrie}
\end{figure*}
In a previous study \cite{bedrunka2021}, we have shown that the collision operator stands out as a suitable solution for integrating \textsc{ml} into the \textsc{lbm}, dynamically adjusting the relaxation rates in response to the evolving characteristics of the flow. In this study, we extend the neural collision operator (\textsc{nco}) $\boldsymbol\Omega:\mathbb{R}^q\to\mathbb{R}^q$ to three-dimensions and design an architecture to ensure its adaptability under equivariance constraints. First, we model $\boldsymbol\Omega$ as follows
\begin{equation}
    \boldsymbol\Omega \left(f_{i}, \theta\right) = - \bm{M}^{-1} S_{\bm{\theta}}(\bm{M}\boldsymbol f - \boldsymbol m^\mathrm{eq}),
    \label{eq:omegatheta}
\end{equation}
where $S_{\bm{\theta}}=\mathrm{diag}(\bm{1}_{\mathrm{dim}+1}, \bm{\omega}_{\nu}, \bm{\omega}_{n}^{\textsc{nn}})$ is a diagonal matrix of relaxation rates. The relaxation rates for conserved moments $\left(\rho, \mathbf{u} \right)$ are set to unity, the ones for the shear moments ($\omega_{\nu}=1/\tau_{\nu}$) are determined by the kinematic viscosity as $\nu = c_s^2 \left(\tau - \frac{\Delta t}{2} \right) $, and the higher-order ones ($\bm{\omega}_{n}^{\textsc{nn}}=1/\bm{\tau}_{n}^{\textsc{nn}}$) are obtained by a neural network $\Phi$. For the D3Q27 stencil, these higher-order moments are grouped into four order groups $n$, each with a determined relaxation rate. The neural network is a function of the transformed moments $\boldsymbol m = \bm{M} \boldsymbol f$ with components $m_i$ or $\boldsymbol m^{eq} = \bm{M} \boldsymbol f^{eq}$  and is trained with respect to local features that depend on the current distribution function component $f_i$.
In this study, the transformation matrix $\bm{M}$ is constructed based on Hermite polynomials. Specifically, the $n$-th order Hermite polynomial, denoted as $\mathcal{H}^{(n)}$, is defined as
\begin{equation}
    \mathcal{H}^{(n)} \left( \mathbf{\xi}_{i} \right) = \dfrac{(-1)^n}{\omega (\mathbf{\xi}_{i})} \nabla_{\mathbf{\xi}_{i}}^n \omega\left(\mathbf{\xi}_{i}\right),
\end{equation}
where $\omega(\mathbf{\xi}_i)$ represents the normalized weight function given by
\begin{equation}
\omega\left(\mathbf{\xi}_{i}\right) = (2 \pi)^{-D/2} \text{exp} \left(-\dfrac{\mathbf{\xi}_{i}^2}{2} \right).
\end{equation}
Additionally, the relationship between the abscisses $\mathbf{\xi}_i$ and the velocity sets $\mathbf{c}_i$ is given by
\begin{equation}
    \mathbf{c}_{i} = {\mathbf{\xi}_i}/{\sqrt{3}}.
\end{equation}
The multivariate nabla operator $\nabla^{(n)}$ provides a compact way to express $n$-th order spatial derivatives. For clarity, it is defined as
\begin{equation}
    \nabla^{(n)} = \nabla^{(n)}_{\alpha_1 ... \alpha_n} = \dfrac{\partial}{\partial x_{a_1}} ... \dfrac{\partial}{\partial x_{a_{n}}}.
\end{equation}
In this study, since we are working in three spatial dimensions ($D=3$), the indices $a_1 \dots a_n$ take values from $\{x, y, z\}$. In order to construct orthogonal vectors $\mathbf{m}$, the transformation matrix $\bm{M}$ is defined as
\begin{equation}
\begin{split}
    \mathbf{M} = (&
    \mathcal{H}^{(0)}_{i},
    \mathcal{H}^{(1)}_{i,x},
    \mathcal{H}^{(1)}_{i,y},
    \mathcal{H}^{(1)}_{i,z},
    \mathcal{H}^{(2)}_{i,xx},
    \mathcal{H}^{(2)}_{i,xy},
    \mathcal{H}^{(2)}_{i,xz},
    \mathcal{H}^{(2)}_{i,yy}, \\
    &\mathcal{H}^{(2)}_{i,yz},
    \mathcal{H}^{(2)}_{i,zz},
    \mathcal{H}^{(3)}_{i,xxy},
    \mathcal{H}^{(3)}_{i,xxz},
    \mathcal{H}^{(3)}_{i,xyy},
    \mathcal{H}^{(3)}_{i,xyz},
    \mathcal{H}^{(3)}_{i,xzz}, \\
    &\mathcal{H}^{(3)}_{i,yyz},
    \mathcal{H}^{(3)}_{i,yzz},
    \mathcal{H}^{(4)}_{i,xxyy},
    \mathcal{H}^{(4)}_{i,xxyz},
    \mathcal{H}^{(4)}_{i,xxzz},
    \mathcal{H}^{(4)}_{i,xyyz}, \\
    &\mathcal{H}^{(4)}_{i,xyzz},
    \mathcal{H}^{(4)}_{i,yyzz},
    \mathcal{H}^{(5)}_{i,xxyyz},
    \mathcal{H}^{(5)}_{i,xxyzz},
    \mathcal{H}^{(5)}_{i,xyyzz},\\
    &\mathcal{H}^{(6)}_{i,xyxzyz} )^T.
\end{split}
\end{equation}

T.S. Cohen and M. Welling \cite{cohen2016,cohen2021} introduced group equivariant convolutional networks to be more robust to symmetry operations and to reduce the required amount of training. The discrete lattice velocities $\left( e_i, \quad i \in \{0,1,\dots q-1\}\right)$ are invariant with respect to a symmetry group $\mathcal{G}$ described by reflections and discrete rotations of $\pi/2$ radians in all directions. These operations span the discrete dihedral group $\mathcal{G}=D_4$ of order 8 and the full octahedral group $\mathcal{G}=O_h$ of order 48 in 3D. Each group operation $g\in\mathcal{G}$ corresponds to a permutation $\Pi[g]$ of discrete lattice velocities locally and thus to a permutation of distribution functions $f_{i}=(f_0,\dots, f_{q-1})$. This permutation can be expressed using permutation matrices $\boldsymbol\Pi_g$ corresponding to each group operation and either acting on the distribution function vector $\boldsymbol f$ or the collision operator $\boldsymbol\Omega$. Exemplarily, Figure \ref{fig:symmetrie} illustrates the group operations applied to the $D2Q9$ stencil, revealing rotation and reflection operations as well as their combinations. It would be a similarly beneficial property if the collision operator be equivariant with respect to $\mathcal{G},$
\begin{equation}
      \boldsymbol\Omega \circ \Pi[g](f_i) =  \Pi[g] \circ \boldsymbol\Omega(f_i) \quad \text{for all}\  g \in \mathcal{G},
\end{equation}
meaning that the collision function $\boldsymbol\Omega$ commutes with the group action. The special form of the \textsc{mrt} model used in this paper was found to be equivariant for both D2Q9 and D3Q27 under the group action. For all possible reflections and rotations, it was checked numerically that equivariance was fulfilled. This required checking that
\begin{equation}
\boldsymbol\Pi_g \boldsymbol M^{-1} S_\theta \boldsymbol M = \boldsymbol M^{-1} S_\theta \boldsymbol M \boldsymbol\Pi_g~.
\end{equation}
The commutation of these matrices could be traced back to ordering the moments according to their order and simultaneously relaxing moments of equal order with the \emph{same} constant relaxation rate. This results in groups of constant elements in the diagonal matrix $S_\theta$. Relaxing this condition, thereby allowing for distinct relaxation rates, immediately violated the equivariance property of $\boldsymbol\Omega$. In order to furthermore guarantee invariance of the neural network with respect to the various group actions discussed above, further steps needed to be taken.
\begin{figure}[h!]
\centering
\includegraphics[height=6cm]{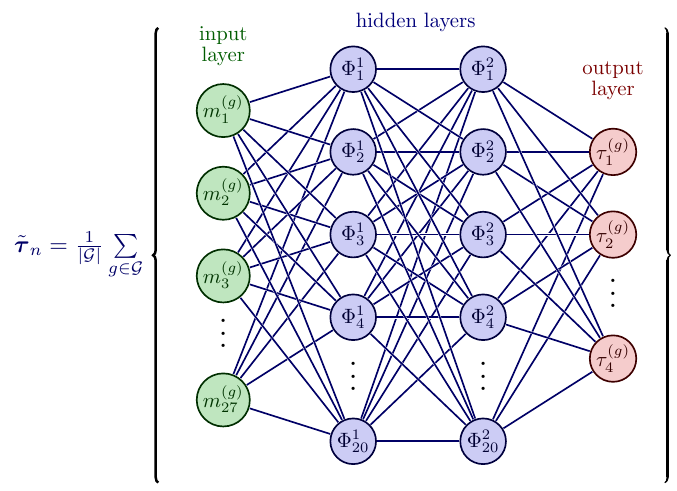}
\caption{Illustration of the neural network topology, including the standard network with two hidden layer and the averaging resulting in an invariant neural network, see eq. (\ref{eq:gconv_m}).}
\label{fig:NN}
\end{figure}

The permutations are used to account for the changing feature inputs when applying the group operations. As stated above, the transformation matrix $\bm{M}$ systematically organizes the moments into groups of their corresponding order. Subsequently, the moments within each group are relaxed collectively, maintaining consistency within their respective order $n$. Each permutation therefore leads to a different ordered set of moments $\boldsymbol m^{(g)} = \boldsymbol M \boldsymbol\Pi_g \boldsymbol f$ as inputs for a standard neural network $\boldsymbol\Phi$, which consists of two layers with 20 neural nodes each and 1064 parameters (see figure \ref{fig:NN}). The network was trained to provide the relaxation rates of the higher-order moments. An invariant neural network $\boldsymbol N_\Phi$ can thus be defined as an average over the networks using all transformed moments $\boldsymbol m^{(g)}$ as inputs
\begin{equation}
    \label{eq:gconv_m}
    \tilde{\bm{\tau}}_{n}: \boldsymbol N_\Phi (\boldsymbol f) = \dfrac{1}{\vert \mathcal{G} \vert} \sum_{g \in \mathcal{G}} \boldsymbol\Phi( \boldsymbol m^{(g)}),
\end{equation}
where $\boldsymbol\Pi_g$ describes the transformation operator of the group $\mathcal{G}$ and $\vert \mathcal{G} \vert$ denotes the number of elements in the group (48 for the full octahedral group in 3D). This averaging ensures that the neural network exhibits group-invariant properties, making it robust to transformations described by $\mathcal{G}$.
This can be seen by simply applying a different transformation $\boldsymbol\Pi_i$ via
\begin{equation}
\boldsymbol N_\Phi(\boldsymbol\Pi_i\boldsymbol f) = \frac{1}{\vert \mathcal{G} \vert} \sum_{g \in \mathcal{G}} \boldsymbol\Phi(\boldsymbol M \boldsymbol\Pi_g \boldsymbol\Pi_i \boldsymbol f ).
\end{equation}
Since the complete octahedral group $\mathcal G$ is closed under multiplication of transformations, the composition of transformations is again an element of the group. Therefore, the neural network is invariant as the sum is including  the same elements $\boldsymbol\Phi(  \boldsymbol m^{(g)})$, just in a different order to lead to:
\begin{equation}
\boldsymbol N_\Phi(\boldsymbol\Pi_i\boldsymbol f) = \boldsymbol N_\Phi(\boldsymbol f).
\end{equation}

\noindent Furthermore, we assert that the learned relaxation parameters are $>0.5$ to ensure over-relaxation and thus stability, by applying a sigmoid function $\sigma$ at the output as
\begin{equation}
    \bm{\tau}_{n} = \dfrac{1}{2(1+e^{ \tilde{\bm{\tau}}_{n}})} + 0.5.
\end{equation}

\subsection{Training Data - Forced Isotropic Turbulence} \label{sec:training}
\begin{figure}[h!]
    \vspace{4cm}
    \textbf{\scriptsize{\hspace{-.2cm}A. \hspace{5.cm} B.}}\par\medskip
    \vspace{-4cm}
    \centering
    \includegraphics[width=0.44\linewidth/1,valign=t]{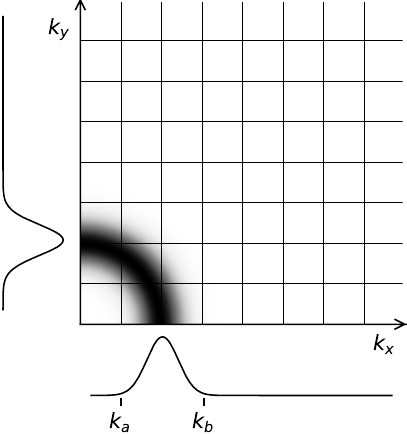}
    \includegraphics[width=0.072\linewidth/1,valign=t]{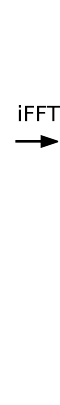}
    \includegraphics[width=0.44\linewidth/1,valign=t]{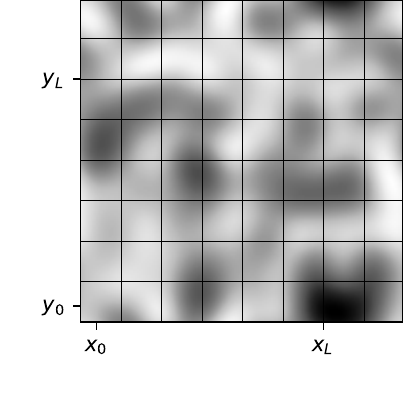}
    \caption{2D representation of the spectral force forcing. \scriptsize\textbf{A}.:\small{} A Gaussian distribution stimulates specific wave numbers in Fourier space. \scriptsize\textbf{B}.:\small{} The force field is then obtained by an inverse Fourier transformation.}
    \label{fig:spectral_force}
\end{figure}
In this study, we employ simulations of forced isotropic turbulence as training data. This flow offers an infinitely variable range of distribution functions due to randomly generated force fields. At the same time, the deliberate injection of energy enables predictable turbulence statistics in the converged state, which serves as the basis for reference as ground truth in the loss function. Kolmogorov's hypothesis of local isotropy due to independence of the small and large scales of homogeneous turbulence is oftentimes invoked. Indeed, local triadic interactions in Fourier space support this hypothesis, leading to a decoupling of the smallest and largest eddies as energy is, in the mean, transported to the dissipative scales. Additionally, a direct coupling between small and large scales within distant triadic interactions is reported in the literature, too, see Brasseur \cite{Brasseur91}, for example. Nevertheless, this additionally motivates us using forced isotropic turbulence as training data to obtain realistic values for higher-order moments within neural networks to obtain a physically based dissipation to stabilize simulations at higher Reynolds numbers, which is furthermore transferable to other flows in the best case.

The effects of various forcing methods for achieving forced isotropic turbulence within the \textsc{lbm} were analyzed. Among these, spectral forcing was found to have distinct advantages, including the ability to generate statistically steady turbulence levels with minimal spurious correlations compared to other established stimulation methods, such as trigonometric forcing. Furthermore, spectral forcing effectively omits compressibility effects by reducing the divergence of the velocity field, which, in turn, decreases numerical errors within the \textsc{lbm} simulations. Following these observations, we utilize Alvelius' \cite{alvelius1999} developed spectral models that inject a power input $\mathcal{P}$ over a discrete time interval $\Delta t$ by stimulating random spherical spheres as shown in Figure \ref{fig:spectral_force}. This method involves a force function $\hat{F}_{\alpha}\left( \mathbf{k,t}\right)$ in a Fourier space that distributes the force across a range of small wave numbers facilitated by randomly generated phases. The force function reads
\begin{equation}
    \hat{F}_{\alpha}\left( \mathbf{k,t}\right) = A_{\text{ran}}\left( \mathbf{k,t}\right)e_{1\alpha}\left( \mathbf{k}\right) + B_{\text{ran}}\left( \mathbf{k,t}\right)e_{2\alpha}\left( \mathbf{k}\right)
\end{equation}
with unit vectors $e_{i\alpha}$ chosen to be normal to the wave number vector $\mathbf{k}$ and orthogonal to each other. This mathematical constraint leads to a divergence-free force field $\mathbf{k} \cdot \hat{\mathbf{F}} =0$ and is satisfied by
\begin{equation}
    e_{1x} = \frac{k_{x}k_{y}}{\sqrt{\left( k^2_x + k^2_y\right)}}, \quad e_{1y} = - \frac{k_{x}k_{y}}{\sqrt{\left( k^2_x + k^2_y\right)}}, \quad e_{1z} = 0,
\end{equation}
and
\begin{equation}
    e_{2x} =\! \frac{k_{x}k_{z}/k}{\sqrt{\left(k^2_x\! + \!k^2_y\right)}}, ~ e_{2y} = \frac{k_{y}k_{z}/k}{\sqrt{\left(k^2_x\! + \!k^2_y\right)}},
    ~e_{2z} = -\frac{\sqrt{\left(k^2_x\! +\! k^2_y\right)}}{k},
\end{equation}
with $k = \vert \mathbf{k} \vert$ \cite{alvelius1999}. The symbols $A_{\text{ran}}$ and $B_{\text{ran}}$ describe random complex numbers, defined as
\begin{equation}
    A_{\text{ran}} = \left( \frac{F\left(k\right)}{2 \pi k^2} \right)^{1/2} \text{exp}\left(i \theta_1 \right)g_{\text{A}} \left( \phi \right),
\end{equation}
\vspace{-.4cm}
\begin{equation}
    B_{\text{ran}} = \left( \frac{F\left(k\right)}{2 \pi k^2} \right)^{1/2} \text{exp}\left(i \theta_2 \right)g_{\text{B}} \left( \phi \right).
\end{equation}
The force distribution is mainly determined by the force density function
\begin{equation}
    F\left(k\right) = A \; \text{exp} \left( - \frac{\left(k-k_{\text{f}}\right)^2}{c}\right)\,,
\end{equation}
shaped in this study by a Gaussian distribution. The symbols $k_{\text{f}}$ and $c$ determine the wave number peak of the forcing spectrum and the width of the Gaussian distribution, respectively. The parameter $A$ describes the amplitude of the forcing depending on the power input $\mathcal{P}$ as
\begin{equation}
    A = \frac{\mathcal{P}}{\Delta t}~ {\bigg/}{\int_{k_{a}}^{k_{b}}\text{exp} \left( - \frac{\left(k-k_{\text{f}}\right)^2}{c}\right)\text{d}k}.
\end{equation}

In his formulation, Alvelius introduced two stochastic variables, $g_{\text{A}}$ and $g_{\text{B}}$, that conform to the constraint $g_{\text{A}}^{2}\left( \phi \right) + g_{\text{B}}^{2}\left( \phi \right) = 1$, with $\phi$ being a uniformly distributed random number within the interval $\left[ 0, \pi\right)$. This random number is generated anew for each wave number and time step. The variables $g_{\text{A}}$ and $g_{\text{B}}$ are determined by the equations
\begin{equation}
    g_{\text{A}} = \text{sin}\left(2\phi\right) \quad \text{and} \quad g_{\text{B}} = \text{cos}\left(2\phi\right)\,,
\end{equation}
ensuring the normalization of these stochastic factors.

The angles $\theta_1$ and $\theta_2$ are related through $\psi = \theta_2 - \theta_1$, where $\psi$ is also a uniformly distributed random number, within the range $\left[ 0, 2\pi\right)$. The magnitude of $\theta_1$ is then calculated using the tangent function,\\[-.7cm]
\begin{multline}
\vspace{-.3cm}
e   \tan \theta_1 =\\
    \frac{\hfill g_{\text{A}}(\phi)\mathrm{Re}\{\xi_1\} + g_{\text{B}}(\phi)(\sin \psi \mathrm{Im}\{\xi_2\} + \cos \psi \mathrm{Re}\{\xi_2\})}{-g_{\text{A}}(\phi)\mathrm{Im}\{\xi_1\} + g_{\text{B}}(\phi)(\sin \psi \mathrm{Re}\{\xi_2\} - \cos \psi \mathrm{Im}\{\xi_2\})},
\end{multline}
where each $\xi_i$ corresponds to the projection of the velocity field along the unit vectors by $\xi_i = \hat{u_\alpha}\text{e}_{i\alpha}$, thereby integrating both the direction and magnitude of the velocity field into the force distribution process. In this study, training data is generated within a cubic domain with an edge length of $2\pi$ on a grid size of $256^3$. These simulations emulate the dynamics of a fluid characterized by a viscosity of $\nu=xx$ and subjected to an energy input rate of $\mathcal{P} = 0.1$.

\subsection{Learning Task}
In accordance with the simulation discussed for forced isotropic turbulence, we generate a series of highly resolved distribution function batches on a grid size of $256^3$ ($f^{\textsc{dns}}_{t=t_0,\dots,t_N}$), which are downsampled to a coarser grid to achieve a distribution function on a grid size of $32^3$. The transformation from fine to coarse is achieved trough acoustic scaling in the form of
\begin{equation}
f_{i,c} = f_{i}^{\text{eq}}\left(\rho \left(\mathbf{x}_{f\rightarrow{}c} \right), \mathbf{u} \left(\mathbf{x}_{f\rightarrow{}c} \right)\right) + \dfrac{2\tau_{c}}{\tau_{f}} f_{i,f}^{\text{neq}} \left(\mathbf{x}_{f\rightarrow{}c} \right),
\label{eq:loss}
\end{equation}
where $f$ and $c$ correspond to the finer and coarser underlying grid, respectively. The non-equilibrium part of the distribution function is defined through the difference between the distribution function and the equilibrium distribution function as $f_{i}^{\text{neq}}=f_{i}-f_{i}^{\text{eq}}$. The dissipation rates of a collision operator correlate with the shape of the resulting energy spectrum. Consequently, the goal is to preserve the true energy spectrum, as obtained from the Direct Numerical Simulation (\textsc{dns}) data. This is achieved by optimizing the loss function:
\begin{equation}
\begin{split}
\!\!\!\!\!L\left(\bm{\theta} \right) = \mathrm{MSE} \left( E(\kappa, \bm{\theta}),{E}(\kappa)\right)+ \omega E(\kappa\!=\!\kappa_{max}),
\end{split}
\quad \text{\!\! } 4 \leq \kappa \leq 10.
\label{eq:loss}
\end{equation}
The loss function compares the shape of the energy spectrum for the wave number range $4 \leq k \leq 10$ between the results obtained using the \textsc{nco} (see eq. \ref{eq:omegatheta}) and the ground truth.
The initial modes are subject to fluctuations due to random spectral excitation and were therefore excluded from consideration. Another important extension of the loss function consists in adding the expression $\omega E(\kappa=\kappa_{max})$. This term, active at the largest wave number, is further minimized to inhibit the accumulation of energy in the higher modes and to tailor the amount of dissipation added for stabilization of under-resolved flows. This minimization is controlled through the weight $\omega$. Consequently, dissipation can be baked into the \textsc{nco} by construction similar to implicit LES modeling and contrary to stabilizing a simulation by means of artificial diffusivity or direct filtering (\cite{Mathew:2006,Fiorina2007,Ricot2009}), although a similar energy spectrum might emerge.

The loss function is optimized with a stochastic gradient descent optimizer (i.e. \textsc{adam}) with a learning rate of 1e-4. It was observed that a higher learning rate led to an unstable training process. The training data set includes 40 batches and is processed in 90 epochs. To incrementally improve stability, the loss function is initially executed after 200 simulation steps on the coarser grid. Subsequently, the number of time steps is increased by 200 time steps every ten epochs. After 90 epochs the simulation completes one full integral time scale, after each application of the loss function.


\section{Training based on forced isotropic turbulence}\label{sec:results}
In this chapter, we explore the calibration process to achieve the desired dissipation behavior. This investigation involves adjusting the weighting factor $\omega$ within the loss function, a parameter that significantly impacts the energy spectrum at the highest wave number. By fine-tuning this factor, we aim to control the dissipation rate in parallel to still ensure that the energy spectrum aligns with theoretical expectations and empirical observations for lower wave numbers. Then, we evaluate an appropriate parameter set across a range of flow scenarios. Specifically, the three-dimensional Taylor-Green vortex is explored, characterized by deterministic flow phases, including laminar and quasi-turbulent flows. Finally, we explore the flow dynamics around a three-dimensional cylinder, offering insights into its capabilities in simulating complex fluid dynamics. The network remains shallow in order to keep the computational costs feasible. In the first application, it consists of two hidden layers with 20 nodes each separated with a ReLU function. However, it is straightforward to expand the network in future studies, making use of a more complex or a deeper network architecture to enhance performance. Here, the emphasis was first on balancing the trade-off between computational cost and model sophistication.

\subsection{Forced Isotropic Turbulence}
\begin{figure}[!b]
    \centering
    \includegraphics[width=0.90\linewidth]{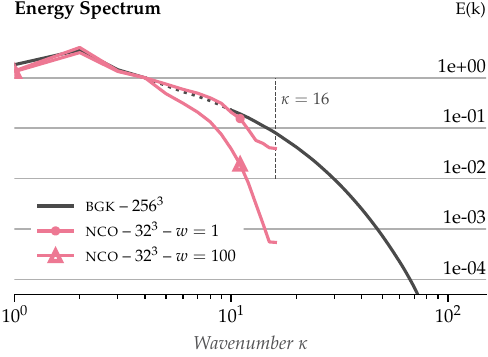}
    \caption{Impact of the weight on the last wave number in the loss function after training: a smaller weight results in energy accumulation at higher scales, whereas a larger weight causes a sharp decline in energy at these scales. The dashed segment of the graph represents the modes with minimized discrepancy relative to the reference.}
    \label{fig:training}
\end{figure}
Figure \ref{fig:training} illustrates the energy spectrum for two different training procedures applied to the \textsc{nco}, which differ primarily in the weighting of the last wave number in the loss function. A lower weight leads to closer agreement with the \textsc{dns} energy spectrum in the wavenumber range. However, the low weight leads only to small amounts of numerical dissipation affecting the highest wavenumbers, leading to unstable simulations in strongly under-resolved simulations. Conversely, a higher weight induces excessive energy dissipation, already affecting the spectrum at low wave numbers.

\begin{figure}[!b]
    \centering
    \includegraphics[width=0.90\linewidth]{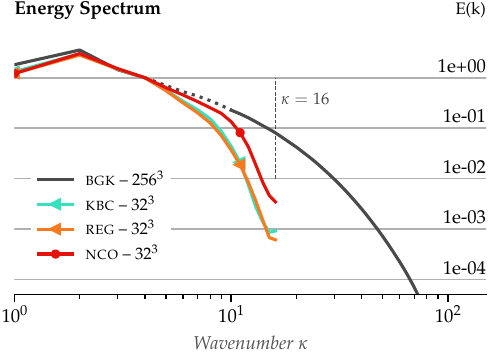}
    \caption{Energy spectrum comparison: analyzing different collision operators relative to \textsc{dns} data with a weight of $\omega=20$ used in the training procedure of the nco. The dashed segment of the graph represents the modes with minimized discrepancy relative to the reference.}
    \label{fig:nco_spectrum}
\end{figure}

Therefore, an intermediate value was chosen to strike the balance between accuracy and robustness. Figure \ref{fig:nco_spectrum} shows the energy spectrum for the trained \textsc{nco} using a weight of $\omega = 20$ to adjust the influence of $E(\kappa\!=\!\kappa_{max})$ within the loss function. To assess this choice of weight on the intrinsic numerical dissipation, the energy spectrum was compared with the results obtained using the Karlin-B\"osch-Chikatamarla (\textsc{kbc}) operator \cite{Karlin2014}, the regularized operator (\textsc{reg}) \cite{Latt2006} and the standard \textsc{BGK} model that provides the \textsc{dns} data. It is obvious that, compared to other commonly used collision operators, the \textsc{nco} maintains the profile of the energy spectrum more accurately before dropping off at higher wave numbers. This choice of weighting parameter leads to a smaller amount of dissipation than other \textsc{lbm} methods, but still enough to guarantee stable behavior, as demonstrated in the upcoming examples. Therefore, the training succeeded in preventing the accumulation of energy at higher wave numbers while at the same time more accurately capturing the energy spectrum of the reference. The late decline of the energy spectrum close to the grid cutoff resembles the situation seen when using approximate deconvolution (\textsc{adm})) or direct filtering instead of explicit \textsc{les} modeling \cite{Stolz:1999,Mathew:2003, Mathew:2006, Sagaut:2006,Ricot2009,Marinc2012}. The filter transfer function cutoff is shifted to wave numbers close to the grid cutoff, thereby minimizing the affected wave numbers due to dissipation.

An other known approach is to use the numerical schemes directly to generate dissipation in implicit \textsc{les} procedures (see, e.g. \cite{Meinke2002,ImplicitLES,Adams2009}). In a similar fashion, the neural collision operator is constructed to include the dissipation required to stabilize the simulations appropriately and, as such, does not necessarily require an explicit \textsc{les} model.

\subsection{Convergence}
To further validate the precision of the \textsc{nco}, we initiate a convergence study utilizing the two-dimensional Taylor-Green vortex (\textsc{tgv2d}). The \textsc{tgv2d}, a canonical problem for examining the numerical accuracy and stability of computational fluid dynamics codes, serves as an ideal benchmark due to the availability of an analytical solution. This study meticulously evaluates the convergence behavior of the \textsc{nco} within the \textsc{lbm} by varying mesh resolutions and assessing its impact on the flow physics comparing it to the theoretical predictions. Specifically, we quantified the deviation of the simulated velocity and pressure field from their theoretical values after one time step. The analytic solution reads\\[-.2cm]
\begin{equation}
\mathbf{u}\left(\mathbf{x}\right) = \left(
\begin{array}{c}
 \hfill U_0 \cos{(x)}\sin{(y)} \\
  - U_0 \sin{(x)}\cos{(y)}\hfill \\
\end{array}%
\right) \exp{}^{-2 \nu t}
\end{equation},
\vspace{-.3cm}
\begin{equation}
p\left(\mathbf{x}\right) = \frac{\rho}{4} \left( \cos{ \left(2x\right)}+\cos{ \left(2y\right)} \right) \exp{}^{-4 \nu t}.
\end{equation}
Given the three-dimensional operational architecture of the \textsc{nco}, the \textsc{tgv2d} is extended and repeated to the third dimension without introducing a gradient, thereby omitting shear stresses and reproducing the profile of the two-dimensional flow.
\begin{figure}[b!]
    \centering
    \includegraphics[width=0.9\linewidth]{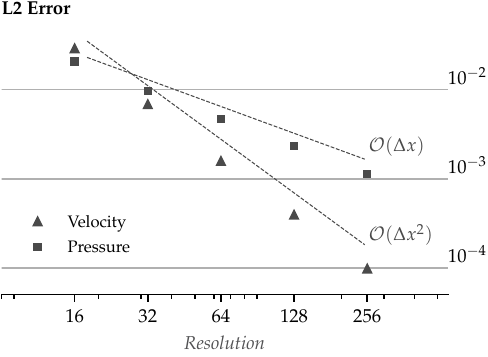}
    \caption{Convergence behavior of the \textsc{nco} within the \textsc{lbm} for the two-dimensional Taylor-Green vortex flow.}
    \label{fig:convergence}
\end{figure}
The findings in Figure \ref{fig:convergence} from this convergence study attest the anticipated reduction in error, as quantified by the L2 norm \cite{kruger2017}.

\subsection{Three-dimensional Taylor-Green Vortex}
\begin{figure}[ht!]
\vspace{-.3cm}
    \centering
    \includegraphics[width=0.32\linewidth]{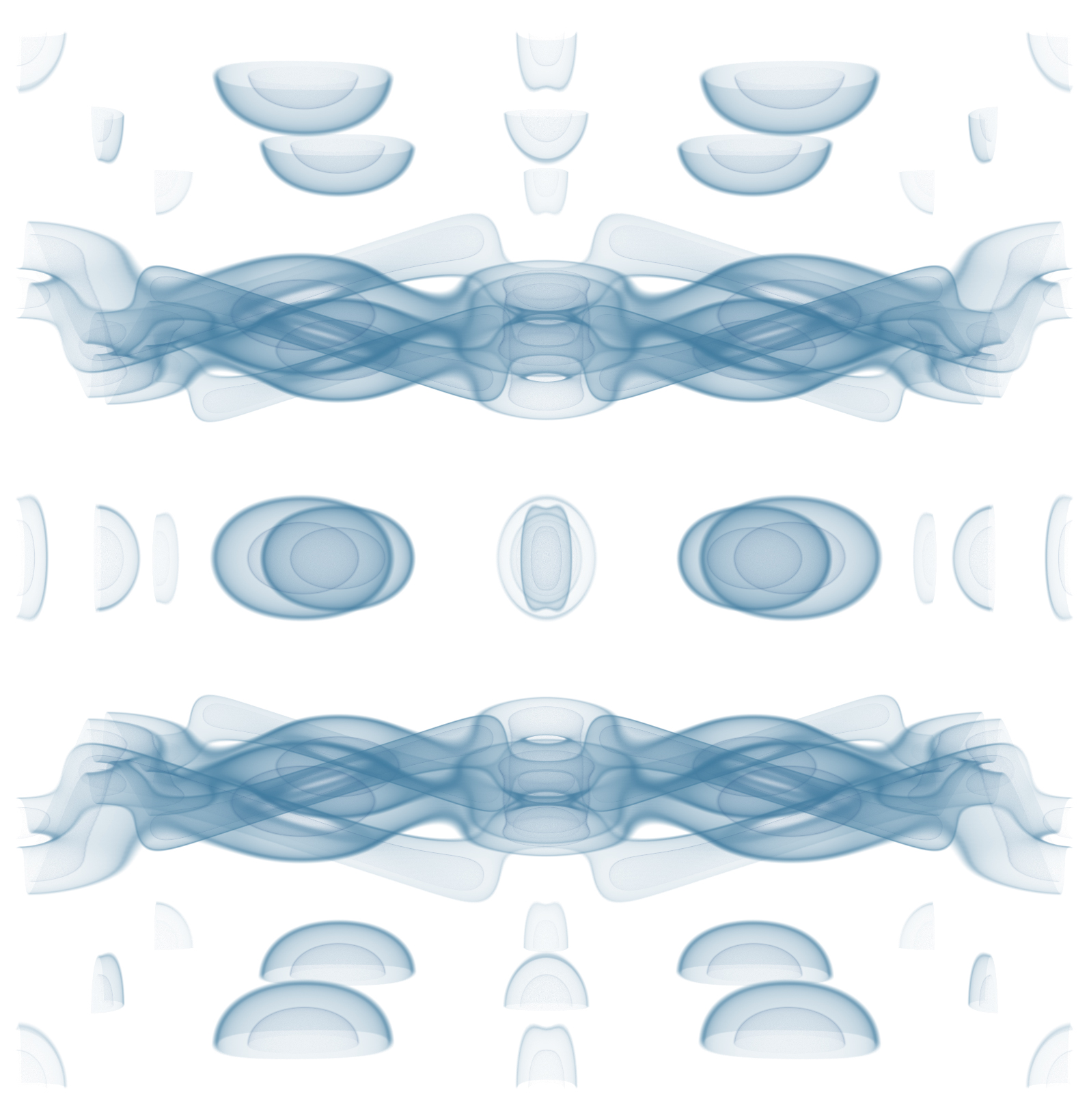}
    \includegraphics[width=0.32\linewidth]{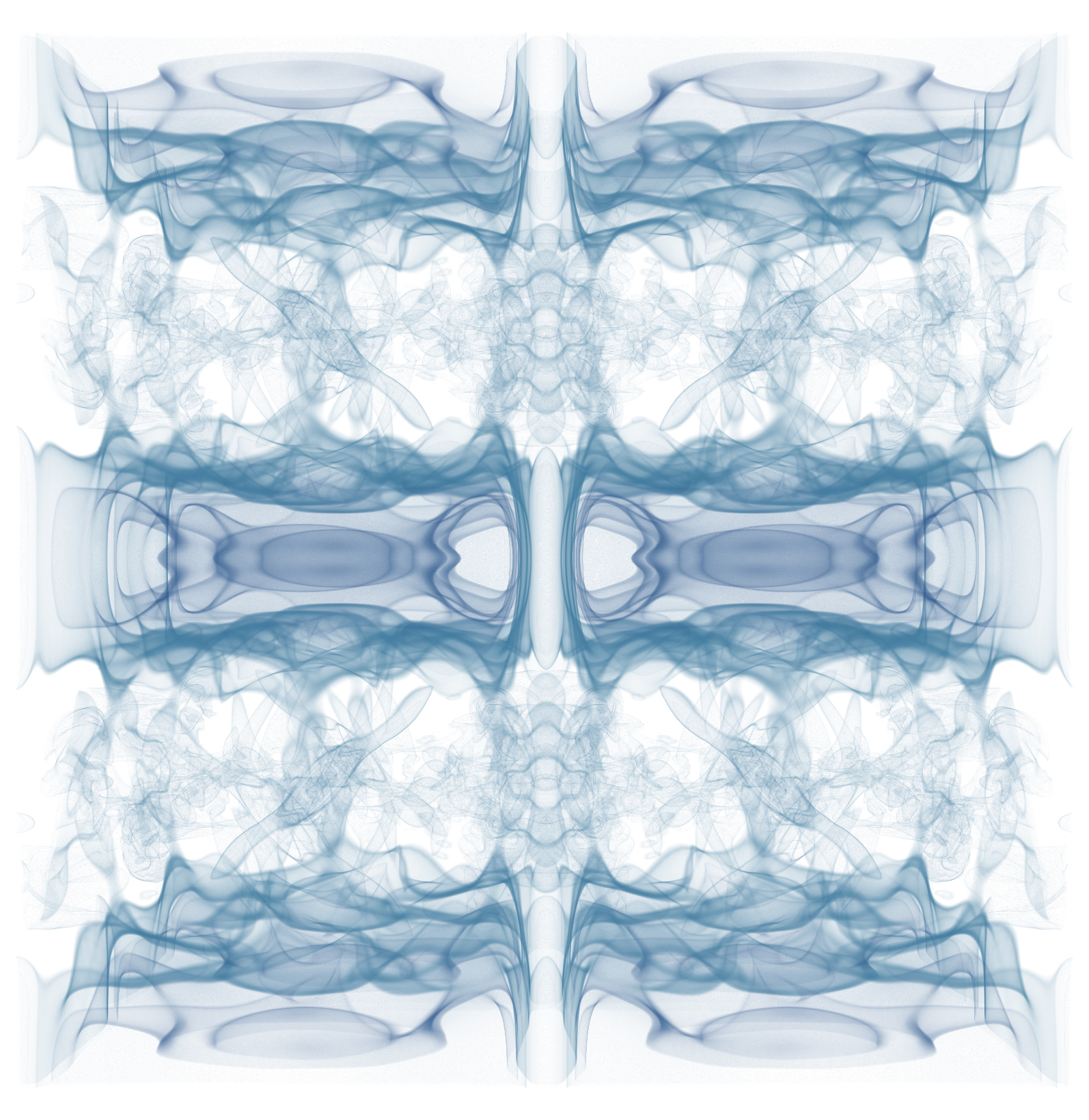}
    \includegraphics[width=0.32\linewidth]{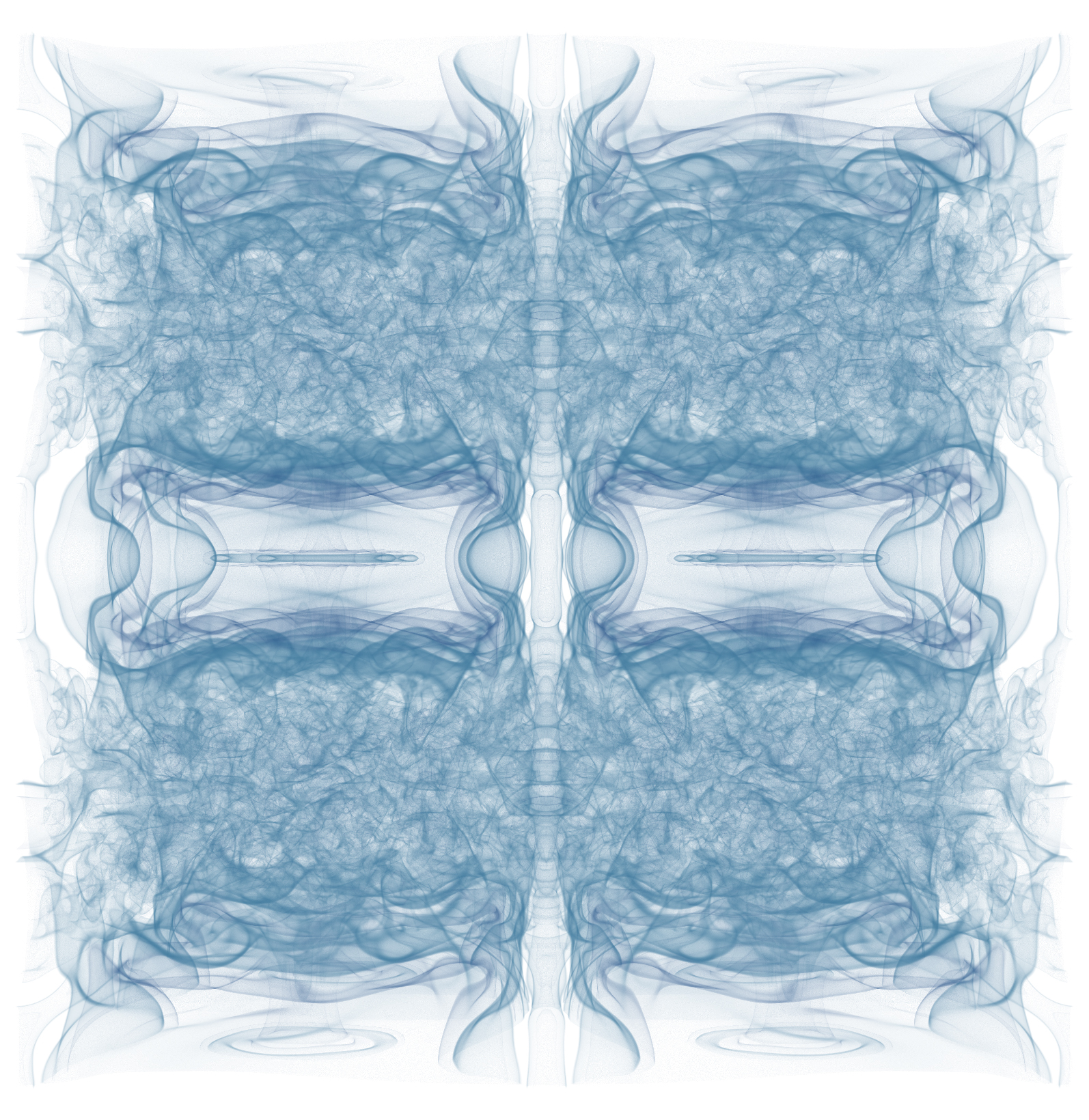}
    \caption{Volumetric rendering of the vorticity of the 3D Taylor–Green vortex at $t=2.5, 7.5, 12.5$, illustrating different deterministic flow phases.}
    \label{fig:tgv3d_vorticiy}
\end{figure}
\begin{figure}[t!]
    \centering
    \includegraphics[width=0.88\linewidth]{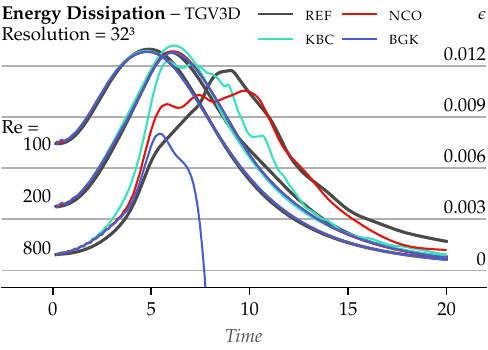}\\
    \vspace{0.2cm}
    \includegraphics[width=0.88\linewidth]{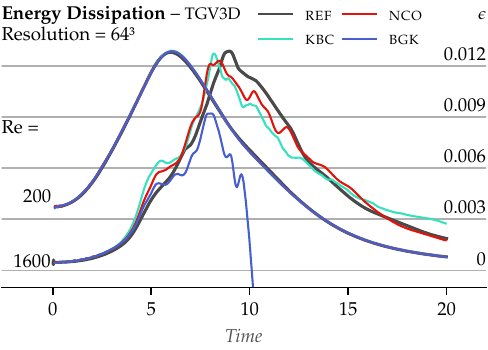}\\
    \vspace{0.2cm}
    \includegraphics[width=0.88\linewidth]{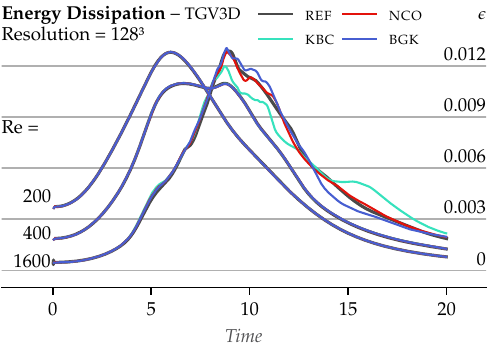}
    \vspace{-.2cm}
    \caption{Energy dissipation $-dk/dt$ of the three-dimensional Taylor-Green vortex for different resolutions ($32^3, 64^3, 128^3$) and Reynolds numbers.}
    \label{fig:tgv3d_dissipation}
\end{figure}

The performance of the trained \textsc{nco} is evaluated simulating a three-dimensional Taylor-Green vortex, providing both a smooth initialized flow field and turbulence throughout the simulations beginning at approximately $t=7.5$ \cite{wang2013} as visible in Figure \ref{fig:tgv3d_vorticiy}. The initial condition on the computational domain $S =\left[ 0, 2\pi \right]$ are
\begin{eqnarray} \label{tgv3d_equation}
\mathbf{u}\left(\mathbf{x},t=0 \right) \!\!\!&=&\!\!\! \left(
\begin{array}{c}
  \hfill \sin{(x)}\cos{(y)}\cos{(z)} \\
  -\cos{(x)}\sin{(y)}\cos{(z)}\\
  0 \\
\end{array}\right),\\
p\left(\mathbf{x},t=0 \right) \!\!\!&=&\!\!\! \frac{1}{16} \left( \cos{(2x)}+\cos{(2y)}\cos{(2z+2)} \right).
\end{eqnarray}
To accurately capture the evolving turbulent profiles after the initialization phase, it is necessary to investigate the associated dissipation rates. As the Reynolds number increases, capturing the correct rates at low resolution becomes increasingly challenging, affecting both accuracy and stability. Figure \ref{fig:tgv3d_dissipation} shows the dissipation rates that are well approximated by $-dk/dt$, due to homogeneity and lack of forcing, for several resolutions and Reynolds numbers. The reference data was simulated at a resolution of $512^3$ grid points using the \textsc{bgk} operator, which is consistent with the findings of Brachet et al. \cite{brachet1983} at a Reynolds number of 1600.
The fundamental \textsc{bgk} operator calculates the predicted profiles with considerable accuracy but lacks stability when the flow field becomes turbulent. In contrast, the \textsc{kbc} operator \cite{Karlin2014} exhibits remarkable stability but fails to accurately capture the reference dissipation rates with higher Reynolds numbers. One can clearly see a strong increase at $t<8$ that provides excessive dissipation. It stands out that the neural collision operator is clearly less dissipative, achieving good agreement with the reference dissipation rate. For $t<8$ the dissipation rate is slightly higher than that of the reference solution but still clearly lower than that of the \textsc{kbc} operator. Thereafter, the peak in the dissipation is slightly underpredicted, without generating instabilities. Therefore, one of the goals of providing more accurate but stable results when becoming strongly under-resolved (visible in the simulation case at a resolution of $32^3$), could be achieved. The \textsc{nco} operator generates even better results than the \textsc{bgk} operator at a resolution of $128^3$ and a Reynolds number of $1600$. The simulations are a first step in demonstrating that using a forced isotropic turbulence simulation as training data for a neural network to determine the higher order relaxations rates is successful and exhibits transferability to other applications.

\subsection{3D Cylinder flow}
\begin{figure*}[ht!]
\vspace{-0.9cm}
\unitlength=1.0truecm
    \centering
    \hspace{-.7cm}\includegraphics[width=0.39\linewidth,angle = -3,clip]{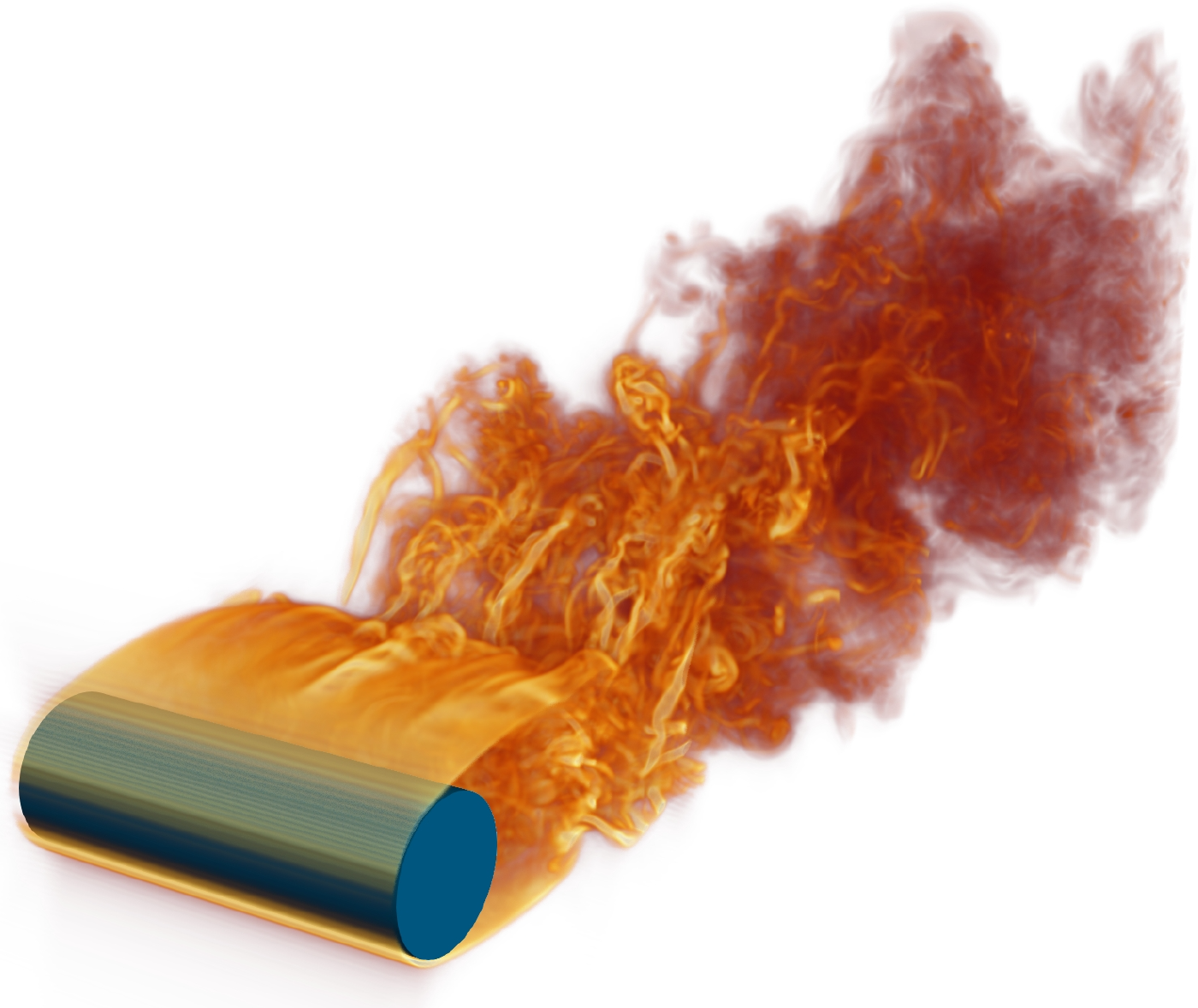}\vspace{-.3cm}
    \hspace{-.5cm}\includegraphics[width=0.57\linewidth]{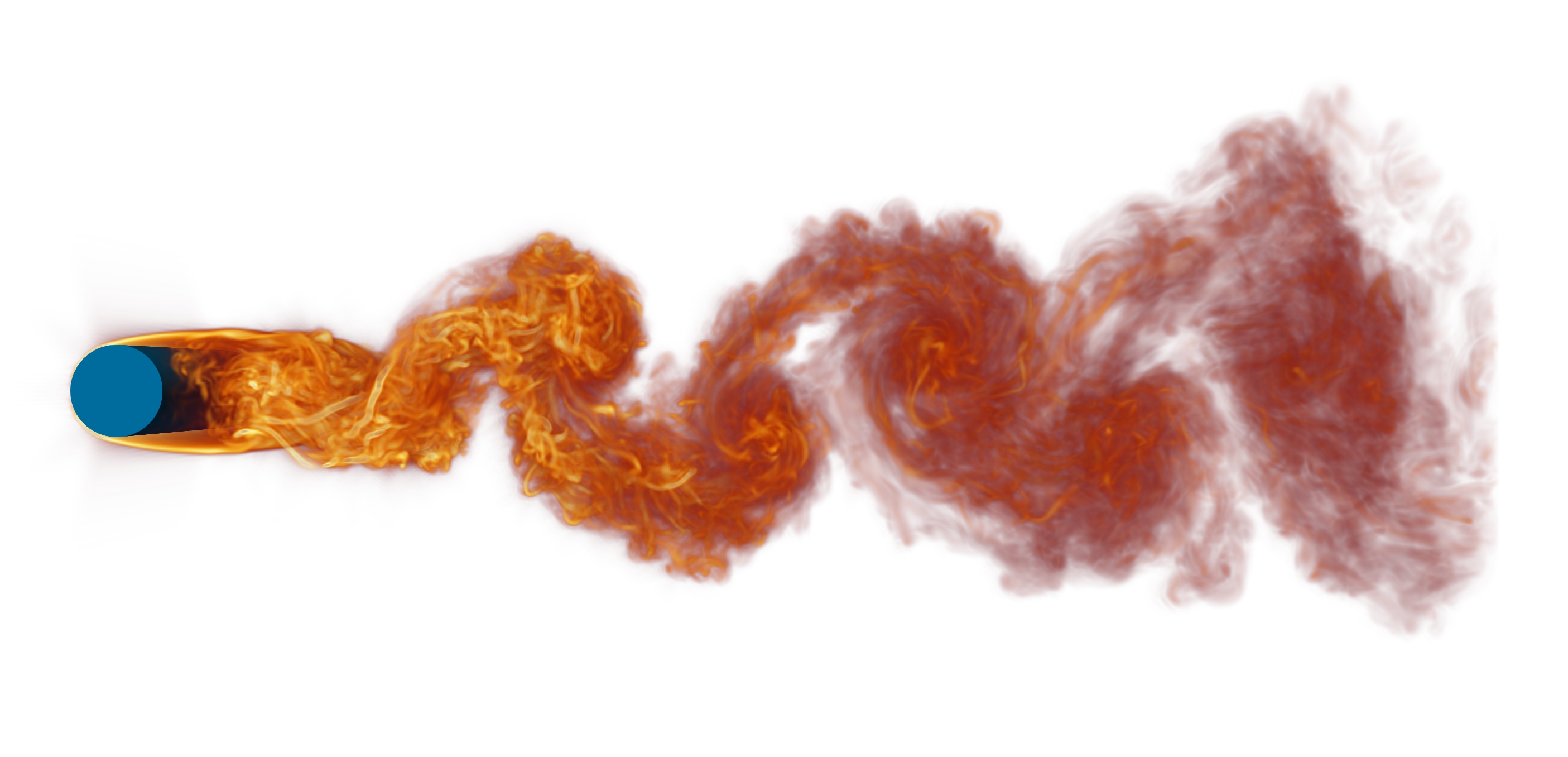}
    \vspace{-.3cm}
    \caption{Volumetric rendering of the vorticity magnitude in three-dimensional cylinder flow, illustrating the K\`{a}rm\`{a}n vortex street phenomenon.}
    \label{fig:Cylinder_Vorticiy}
\end{figure*}

\begin{figure}[!hb]
\vspace{-.1cm}
    \centering
    \includegraphics[width=0.8\linewidth]{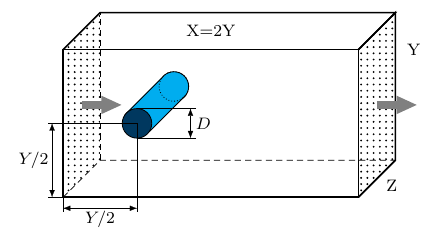}
    \vspace{-.4cm}
    \caption{Domain geometry: The stream-wise domain length X is twice the domain height Y. The cylinder of diameter D is placed at Y/2 from the inlet. Boundary conditions: Inlet on the left, outlet on the right, periodic boundaries for all lateral boundaries.}
    \label{fig:CylinderDomain}
    \vspace{-.5cm}
\end{figure}

The flow around a circular cylinder positioned perpendicularly to the flow direction serves as a classical test case for the fluid-structure interaction. This scenario has been the subject of extensive experimental and simulation studies over decades, yet it remains a popular choice due to its simplicity, relevance, and applicability to more complex scenarios. Therefore, the evaluation of the three-dimensional cylinder flow investigates the practicality of using the \textsc{nco} for real-world applications. The geometry is depicted in Figure \ref{fig:CylinderDomain}. The streamwise length $X$ of the domain is twice the height $Y$, the crosswise length is indicated by $Z$, and the resolution of the lengths is defined relative to the resolution of the diameter $D$.

With the exception of the inlet and outlet regions, the boundaries of the simulation domain are configured to enforce periodic conditions: populations that leave the computational domain are assigned the values of the corresponding populations entering the domain on the opposite side. \\

\textit{Inlet:} the equilibrium boundary condition stands out as the simplest form of a nonperiodic boundary condition, where the particle populations are initialized according to the equilibrium distribution function (Eq. \ref{Eq:EquilibriumDistribution}). Although not fully capturing the complexities of physical interactions within the flow, this method is effectively utilized as an inlet condition. By prescribing the flow's velocity and density at the boundary, it allows for the simulation of inflow without the need for detailed information about the flow's behavior in the bulk of the domain. \\

\textit{Outlet:} An advanced method for simulating open boundaries is the anti-bounce-back boundary condition. This method approximates the behavior of an open fluid volume by extending the properties of the flow into the boundary, thereby allowing for a seamless transition of the fluid properties at the interface. The fluid velocity at the boundary is calculated from the interior fluid region according to
\begin{equation}\label{eq:Anti-bounce-back_BC}
    f_{\bar{i}}(x_b,t+\Delta t)=-f_{i}^{\star}(x_b,t)+2w_i \rho_w \left(1+\frac{(\boldsymbol{c}_i\cdot \boldsymbol{u}_w)^2}{2c_s^4}-\frac{\boldsymbol{u}_w^2}{2c_s^2}\right),
\end{equation}
$x_b$ is the fluid node interacting with the boundary, which lies half a lattice spacing away from it. The velocity $\boldsymbol{u}_w$ can be determined as:
\begin{equation}
    \boldsymbol{u}_w=\boldsymbol{u}(x_b)+\left[\boldsymbol{u}(x_b)-\boldsymbol{u}(x_{b+1})\right]/2,
\end{equation}
where $x_{b+1}$ is the next fluid node following the inward normal vector of the boundary. With anti-bounce-back boundary conditions, constant pressure outlets can be constructed. \\

\textit{Solid:} The interpolated bounce-back boundary method represents an improvement over the simple rudimentary bounce-back method by integrating the precise position of the boundary, denoted as $d$ (i.e. the distance from the fluid node to the solid wall)), into the established bounce-back scheme. This approach utilizes interpolation of bounced populations based on the boundary's relative position, employing either two (linear interpolation) or three (quadratic interpolation) adjacent fluid nodes. The expression for linear interpolation is formalized as:
\begin{equation}
    f_{\bar{i}}(\boldsymbol{x},t+\Delta t)=
    \begin{cases}
        2df_{i}^{\star}(\boldsymbol{x}_{b},t)+(1-2d)f_{i}^{\star}(\boldsymbol{x}_{f},t) \quad & d \le\frac{1}{2}\\
        \frac{1}{2d}f_{i}^{\star}(\boldsymbol{x}_{b},t)+\frac{2d-1}{2d}f_{\bar{i}}^{\star}(\boldsymbol{x}_{b},t) & d >\frac{1}{2}
    \end{cases},
\end{equation}
where $\boldsymbol{x}_{b}$ and $\boldsymbol{x}_{f}$ represent the fluid node directly adjacent to the solid boundary and the second neighboring fluid node, respectively.\\

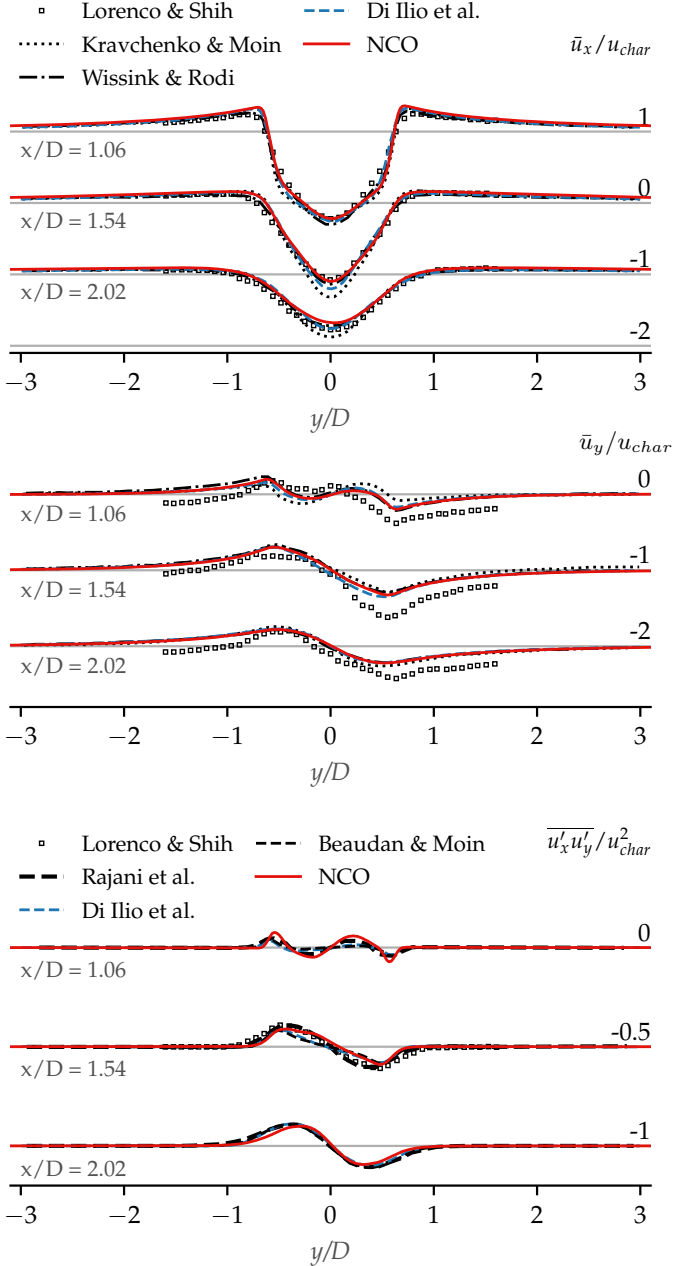
\begin{figure}[ht!]
    \centering
     \vspace{-.0cm}\includegraphics[width=1.00\linewidth]{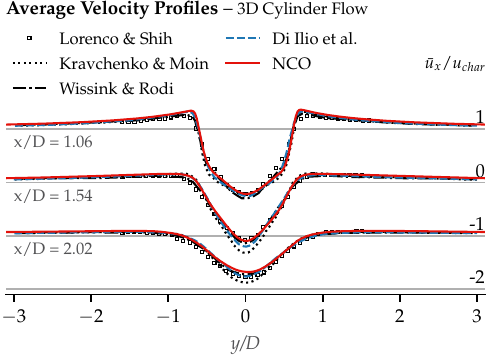}\\
    \vspace{0.1cm}
    \includegraphics[width=1.00\linewidth]{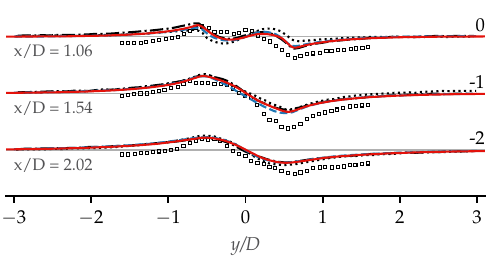}
    \begin{picture}(3,3)
    \put(3.4cm,4.9cm){$\bar{u}_y/u_{char}$}
    \end{picture}
    \vspace{-.4cm}
    \includegraphics[width=1.00\linewidth]{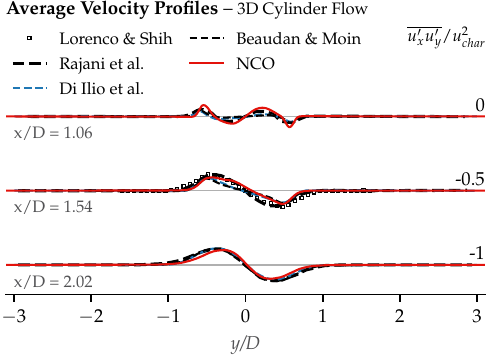}
    \caption{Cylinder flow: Time-averaged  profiles along y-axis for velocities $u_x$ (\textit{top}), $u_y$ (\textit{middle}) and Reynolds stress profiles $\overline{u'_x u'_y}/u^2_{char}$ (\textit{bottom}), complemented by simulations and measurements depicted at different positions along the x-axis. $u_{char}$ denotes the inflow velocity.}
     \label{fig:Cylinder_profiles}
     \vspace{-.7cm}
\end{figure}

In the context of flows around objects such as cylinders, the determination of force coefficients (e.g., drag and lift coefficients) serves as key indicators of the flow behavior and interaction with the object. Therefore, these coefficients are crucial for benchmarking simulations of the cylinder flow. The drag coefficient quantifies the resistance offered by the object against the flow direction, whereas the lift coefficient measures the force perpendicular to the flow direction.

These coefficients are derived from the drag force $F_x$ and lift force $F_y$, respectively, and are calculated by
\begin{equation}\label{eq:drag_lift}
    C_D=\frac{2F_x}{\rho U^2A} \quad \text{and} \quad C_L=\frac{2F_y}{\rho U^2A}
\end{equation}
where $\rho$ is the fluid density, $U$ is the free-stream velocity, and $A$ is the reference area perpendicular to the characteristic direction of the force coefficients. Within the \textsc{lbm}, the momentum exchange method is widely recognized for its straightforwardness and effectiveness in computing the force exerted on an object by the flow by\\[-.3cm]
\begin{equation}
    \boldsymbol{F}={\Delta P}/{\Delta t}.
\end{equation}
The momentum change $\Delta P$ is calculated by aggregating the momentum contributions from incoming$f_i^{in}$ and reflected $f_i^{out}$ distributions at the boundary, expressed as
\begin{equation}\label{eq:MEM_basic}
    \Delta P=\Delta x^3\sum_{x_i^b}\Delta p(\boldsymbol{x}_i^b)=\Delta x^3\sum_{x_i^b}(f_i^{in}\boldsymbol{c}_i-f_{\Bar{i}}^{out}\boldsymbol{c}_{\Bar{i}}).
    \vspace{-.3cm}
\end{equation}
The Strouhal number, derived from the shedding frequency, can be determined by observing the K\`{a}rm\`{a}n vortex street. This characteristic wake pattern is illustrated in Figure \ref{fig:Cylinder_Vorticiy}, which depicts the volumetric vorticity. The simulation performed at a Reynolds number of $Re=200$ and a Mach number of $Ma=0.1$ involves a domain setup characterized by a grid nodes per diameter (\textsc{gpd}) of 30, resulting in a resolution of $x=40\times \textsc{gpd}$, $y=20\times \textsc{gpd}$, and $z=2\times \textsc{gpd}$. The study's evaluation led to drag and lift coefficients of $C_D=1.36$ and $C_L=0.68$, respectively. The simulation demonstrates a good agreement with referenced benchmarks \cite{choi2007, braza1986, russell2003, shu2007, liu1998, wright2001, wright2001}, highlighting the operator's effectiveness.

An increase in the Reynolds number significantly amplifies the magnitudes of turbulence. To facilitate a more detailed analysis, time-averaged velocity profiles are evaluated in the flow's converged state at a Reynolds number of $3900$ illustrated in Figure \ref{fig:Cylinder_Vorticiy}. The dimensions of the setup change to: $\textsc{gpd}=46$, $x=20\times \textsc{gpd}$, $y=10\times \textsc{gpd}$, and $z=3\times \textsc{gpd}$.

Figure \ref{fig:Cylinder_profiles} 
presents the time-averaged velocity and Reynolds shear stress profiles along the y-axis at three different streamwise positions, comparing them to reference studies. These references are derived from experiments (i.e., Lorenceo \& Shih \cite{lourenco1993}) or simulations. Common to the simulations is the use of locally refined grid domains, which enhance the accuracy in capturing peak velocity magnitudes. However, even without additional grid refinement, the \textsc{nco} operator is capable of capturing the essential characteristics and maintains stability throughout the simulation.

\section{Training based on dissipation rates}
\begin{table}[t!]
\centering
\caption{Training Cases: Each combination of Reynolds number, Mach number, and resolution is executed within an epoch.}
\label{tab:tgv_training_combination}
{\small
\begin{tabular}{>{\centering\arraybackslash}m{2cm}%
                >{\centering\arraybackslash}m{1cm}%
                >{\centering\arraybackslash}m{1cm}%
                >{\centering\arraybackslash}m{1cm}%
                >{\centering\arraybackslash}m{1cm}}
\toprule
\makecell{\textbf{Reynolds}\\\textbf{number}} & \multicolumn{4}{c}{\textbf{Mach number}} \\
\cmidrule(lr){2-5}
 & 0.05 & 0.1 & 0.15 & 0.2 \\
\midrule
400 & $16^{3}$ & \makecell[t]{$10^{3}$\\$16^{3}$\\$25^{3}$} & \makecell[t]{$10^{3}$\\$16^{3}$\\$25^{3}$\\$32^{3}$} & \makecell[t]{$10^{3}$\\$16^{3}$\\$25^{3}$\\$32^{3}$} \\
\midrule
800 & $16^{3}$ & \makecell[t]{$10^{3}$\\$16^{3}$\\$25^{3}$} & \makecell[t]{$10^{3}$\\$16^{3}$\\$25^{3}$\\$32^{3}$} & \makecell[t]{$10^{3}$\\$16^{3}$\\$25^{3}$\\$32^{3}$} \\
\midrule
1600 & $16^{3}$ & \makecell[t]{$10^{3}$\\$16^{3}$\\$25^{3}$} & \makecell[t]{$10^{3}$\\$16^{3}$\\$25^{3}$\\$32^{3}$} & \makecell[t]{$10^{3}$\\$16^{3}$\\$25^{3}$\\$32^{3}$} \\
\bottomrule
\end{tabular}}
\end{table}

\begin{figure*}[!ht]
    \centering
    \begin{minipage}{0.32\textwidth}
        \includegraphics[width=\linewidth]{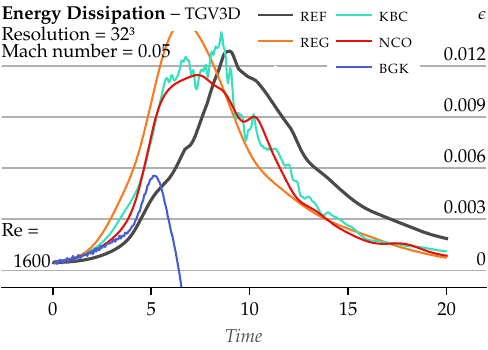}
    \end{minipage}
    \hfill
    \begin{minipage}{0.32\textwidth}
        \includegraphics[width=\linewidth]{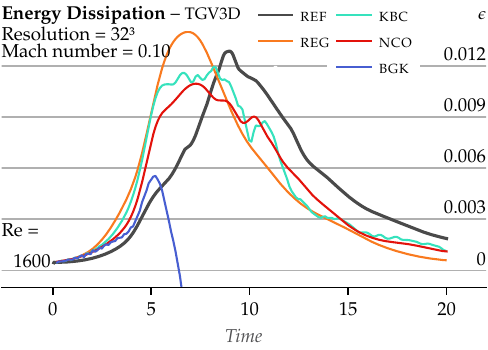}
    \end{minipage}
    \hfill
    \begin{minipage}{0.32\textwidth}
        \includegraphics[width=\linewidth]{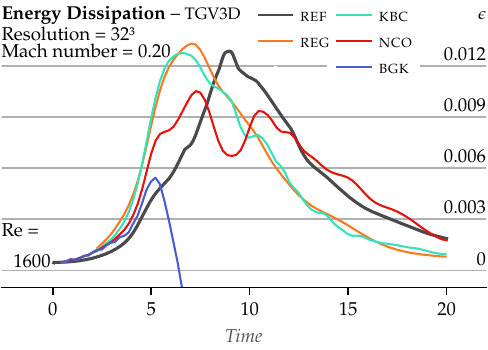}
    \end{minipage}

    \vskip\baselineskip

    \begin{minipage}{0.32\textwidth}
        \includegraphics[width=\linewidth]{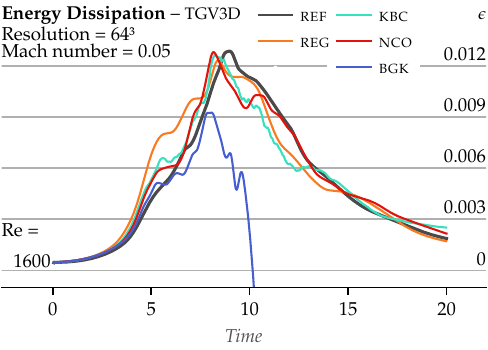}
    \end{minipage}
    \hfill
    \begin{minipage}{0.32\textwidth}
        \includegraphics[width=\linewidth]{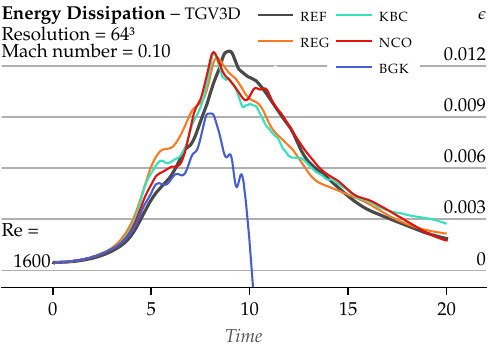}
    \end{minipage}
    \hfill
    \begin{minipage}{0.32\textwidth}
        \includegraphics[width=\linewidth]{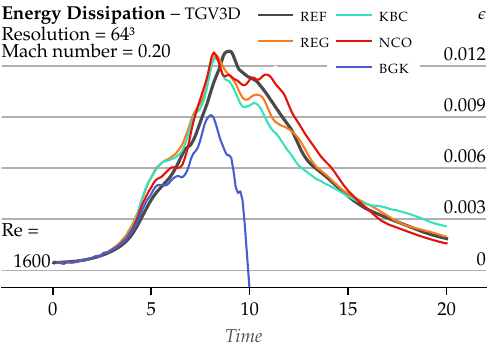}
    \end{minipage}

    \vskip\baselineskip

    \begin{minipage}{0.32\textwidth}
        \includegraphics[width=\linewidth]{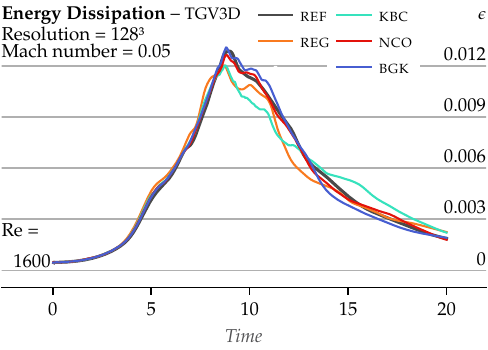}
    \end{minipage}
    \hfill
    \begin{minipage}{0.32\textwidth}
        \includegraphics[width=\linewidth]{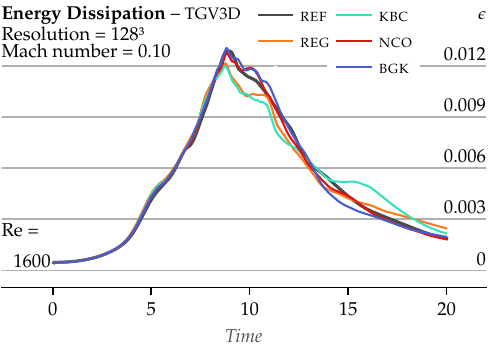}
    \end{minipage}
    \hfill
    \begin{minipage}{0.32\textwidth}
        \includegraphics[width=\linewidth]{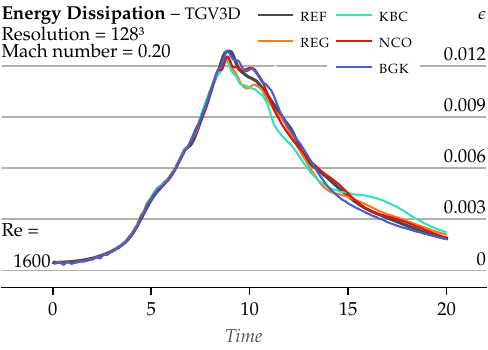}
    \end{minipage}

    \caption{Energy dissipation rate, $-dk/dt$, of the three-dimensional Taylor-Green vortex across various configurations. Rows represent different spatial resolutions, while columns correspond to varying Mach numbers.}
    \label{fig:bilder3x3}
\end{figure*}

Another possibility of optimizing the relaxation rates is to use time-dependent quantities directly. The present section will briefly show the potential of this alternative in determining the relaxation rates of non-physical moments. As a building block, the three-dimensional \textsc{tgv} was chosen, making use of its deterministic nature to train neural networks based on key quantities that change over time. In numerical fluid dynamics, dissipation is often used for benchmarking the solver's accuracy for various Reynolds numbers, as presented above. Therefore, one idea is to use its temporal evolution itself to train for the desired collision operator behavior. Since a training step of the neural network spans multiple simulation steps and the dissipation rate is a global quantity based on local information, a high memory requirement is needed to store the gradients. At this point, one can make use of the symmetries inherent to \textsc{tgv} flow. Since the initial symmetry of the flow is preserved over time, this symmetry can be exploited by quartering the domain along an edge in combination with suitable symmetry boundary conditions. In three dimensions, this results in a memory reduction of a factor of 64. Assuming fixed hardware resources, this allows for a higher grid resolution and enables the simulation of higher Reynolds numbers. The domain truncation along the symmetry planes requires specific boundary conditions to be applied at the new boundary faces - for details, see \ref{Appendix_TGV}.

\subsection{Incompressible kinetic energy}
The configurations described in Table \ref{tab:tgv_training_combination} represent the different combinations of Reynolds numbers, Mach numbers, and resolutions that are executed within each epoch during training. The learning rate used for the training is set at $1\times10^{-4}$. An additional hidden layer containing 20 nodes is introduced to enhance the capabilities of the neural network. The training spans a total of 100 epochs. The loss function used is the mean squared error of the difference in the dissipation rates, $-dk/dt$, between the reference results and the results simulated using the \textsc{nco}. The dissipation is compared across 100 discrete points between $t=0$ and $t=15$. In contrast to training using the energy spectrum provided by forced isotropic turbulence, no extra effort was used here to tailor the dissipation behavior close to the grid cutoff. \\

Figure \ref{fig:bilder3x3} compares the energy dissipation rates obtained using the \textsc{bgk} operator, the \textsc{kbc} operator, the \textsc{reg} operator, and the newly trained \textsc{nco} using time-dependent dissipation data. The results show that the \textsc{nco} not only accurately captures the dissipation behavior, closely aligning with the reference data throughout the time interval at a resolution of $128^3$, but also maintains stability when the resolution is decreased. In contrast, the \textsc{bgk} operator fails to capture turbulent behavior effectively at lower resolutions. The \textsc{kbc} and \textsc{reg} operators demonstrate better stability than \textsc{bgk}, particularly in under-resolved conditions. The \textsc{reg} model introduces excessive dissipation at earlier times at a resolution of $32^3$ grid points over all Mach numbers. Even at a resolution of $64^3$ the \textsc{kbc} and \textsc{reg} operators required slightly higher dissipation at the beginning to maintain stability compared to the \textsc{nco}. The \textsc{kbc} and \textsc{nco} are similar at low Mach numbers, with the \textsc{kbc} model producing higher dissipation rates when the Mach number increases. All three models underestimate the dissipation after reaching the peak value. Only the \textsc{nco} is able to improve for the highest Mach number case. Increasing the resolution to $64^3$ grid points improves the models with the exception of the \textsc{bgk} model. The regularized \textsc{lbm} still produces higher dissipation rates during the transition phase and underestimates it after reaching the peak value for the two lowest Mach number cases. The \textsc{kbc} model is similar to the \textsc{nco} model for early times, but deviates for larger times after reaching the dissipation peak. At the largest Mach number, now the \textsc{nco} produces slightly higher dissipation levels after $t>10$.
The \textsc{nco} demonstrates accuracy in capturing the dissipation behavior, aligning closely with the reference data across all Mach numbers tested at a resolution of $128^3$. The overall consistency across all Mach and resolution numbers tested here highlights the \textsc{nco}’s ability to produce accurate and robust results, a performance that is not achieved similarly by the \textsc{kbc} or regularized (\textsc{reg}) operators.

Overall, the section demonstrated that the reduced domain setup preserved the original dynamics of the Taylor-Green vortex initialization while enabling training at higher Reynolds numbers, which is crucial for turbulent flow simulations. The results indicate that the \textsc{nco} trained on time-dependent dissipation data produce both stable and accurate outcomes. This was achieved even in challenging under-resolved conditions, without attempting to further adjust the numerical dissipation.

\begin{figure}[b!]
    \centering
    \includegraphics[width=1.00\linewidth]{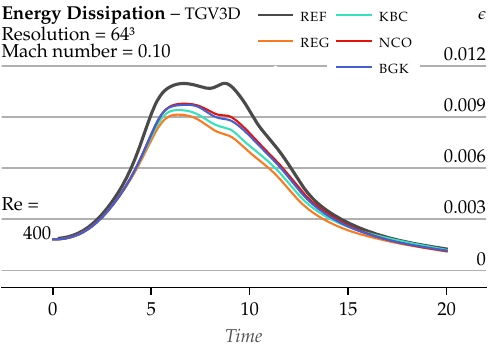}
    \includegraphics[width=1.00\linewidth]{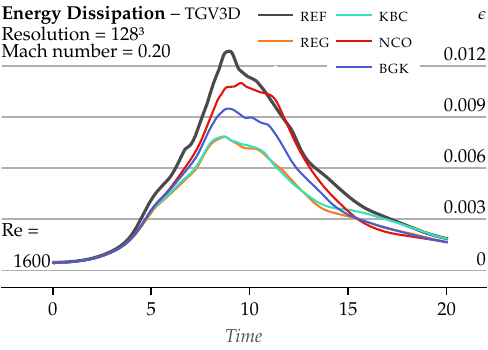}
    \caption{Comparison of energy dissipation rates for different collision operators based on shear rates.}
    \label{fig:tgv_shear_rate}
\end{figure}

\subsection{Shear rates}
While the procedure described previously results in a stable and accurate model, this section examines an alternative training approach for the \textsc{nco}, in which dissipation rates are calculated using shear rates instead of relying on kinetic energy. This method leads to a larger discrepancy between the predicted dissipation and the ground truth, particularly in under-resolved simulations. In general, a much higher resolution is required to capture the right dissipation rates by means of the shear rates. By tuning the relaxation rates based on this procedure, the neural collision operator aims to achieve greater accuracy tackling the dissipation rates directly compared to the other common operators. As shown in Figure \ref{fig:tgv_shear_rate}, the shear rate-based  \textsc{nco} outperformed even the \textsc{bgk} operator in terms of capturing the correct dissipation rates. Despite its high accuracy, the collision operator trained with this shear rate approach exhibited stability issues when applied to strongly under-resolved scenarios or high Reynolds number flows due to insufficient numerical dissipation. In summary, the shear rate-based training approach resulted in an exceptionally accurate \textsc{nco}, particularly suited to capture the correct dissipation behavior in turbulent flow fields. However, the increased accuracy came at the cost of reduced stability, especially in under-resolved simulations. Future work could focus on addressing these stability issues, possibly by incorporating additional regularization during training or combining the shear rate approach with an adaptive stability mechanism. Compared to the more stable but less accurate training discussed, this approach demonstrates the potential to adaptively tune relaxation rates for a specific aim.


\section{Conclusion} \label{sec:conclusion}
This paper describes the development of a neural collision operator \textsc{nco} in three dimensions, based on an invariant network architecture within an equivariant collision operator. Several different ways to determine the relaxation parameters were discussed. In a first step, a simple training procedure was introduced, allowing the fine-tuning of the numerical dissipation to provide accurate and stable results. The neural network parameters were optimized using data from a forced isotropic turbulence simulation which was stimulated by spectral forcing. The training process was designed to reproduce the energy spectrum observed in \textsc{dns} by tuning the higher-order non-physical moments over a maximum range of wave numbers. This training was performed for various resolutions including highly under-resolved simulation data. Furthermore, robustness to guarantee stable simulations was part of the design, by adjusting the dissipation for the wave numbers close to the grid cutoff. Therefore, contrary to simulations using explicit filtering or explicit subgrid models, dissipation is implicitly enforced using the \textsc{nco}, similarly to implicit LES. A tradeoff was chosen between more dissipative models like \textsc{kbc} and the standard but often unstable \textsc{bgk} collision operator. It was demonstrated that the suggested modified loss function provided the ability to adjust the spectrum shape accordingly, with dissipation acting predominantly only at the highest wave numbers. Therefore, the present \textsc{nco} is the result of a well-chosen balance between accuracy and robustness. \\

Upon completion of the training, the \textsc{nco} was assessed through various test cases. A convergence study confirmed the expected behavior of the operator. Moreover, the simulation of the three-dimensional Taylor-Green vortex not only exhibited stable performance at under-resolved conditions and at high Reynolds numbers, but it also closely aligned with reference data showing superior behavior compared to alternative \textsc{lbm} models. In addition, we simulated a three-dimensional cylinder flow as a more complex test case. Here, we accurately computed characteristic physical coefficients, such as drag and lift. Furthermore, the velocity profiles behind the cylinder showed excellent agreement with reference data, thus confirming the model's capability to capture essential physical properties in under-resolved simulations. Supplementing the procedure above, alternative ways of determining the relaxation rates were presented. These consisted of training the \textsc{nco} using time-evolving dissipation rates from the Taylor-Green vortex (\textsc{tgv}). For this, the three-dimensional \textsc{tgv} constitutes an ideal flow case due to its deterministic nature and inherent symmetries. These symmetries in association with specific boundary conditions could be used to reduce the required domain size to achieve training at higher Reynolds numbers with manageable memory requirements. Contrary to basing the training on the dissipation determined using the shear rates directly, this approach proved to be very successful in accurately capturing the dissipation rates. This could be achieved using the constructed \textsc{nco} without sacrificing stability, offering an alternative to train neural networks in computational fluid dynamics.

\section*{Acknowledgments}
We would like to express our sincere thanks to Andreas Krämer and Dominik Wilde for their meticulous inputs, which significantly enhanced the quality of our work. The simulations were performed using the Platform for Scientific Computing at Bonn-Rhein-Sieg University of Applied Sciences, which is funded by the German Ministry of Education and Research and the Ministry for Culture and Science North Rhine-Westphalia.

\appendix
\section{Reduced Taylor-Green vortex} \label{Appendix_TGV}
In this appendix the mathematical derivation of the symmetry boundary conditions is presented. These are used to reduce the simulation domain of the Taylor-Green vortex (\textsc{tgv}), allowing the simulation of only a fraction of the domain, by taking advantage of periodic repetitions in the flow structure. Initially, the trigonometric definition remains consistent with Eq. \ref{tgv3d_equation}; however, it is now defined in the interval $\left[ 0 + \frac{\Delta x}{2}, \frac{\pi}{2} - \frac{\Delta x}{2} \right]$. The symmetry boundary conditions involve transforming the distribution function at the boundaries in each time step to accurately account for symmetric interactions. By leveraging the inherent periodic computation of the algorithm, the outward-pointing distribution vectors $d_{i,\alpha}$ at the boundaries are shifted so that they act on the correct nodes prescribed by the symmetry conditions. The computational domain is discretized so that the nodes are positioned at a distance of $\Delta x / 2$ from the symmetry edges or planes, ensuring no nodes lie directly on the boundaries and allowing for a symmetric discretization of the domain. In this context, the boundary positions are characterized by $x^{-}$ and $x^{+}$, with $x^{-}=\Delta x / 2$ and $x^{+}=\pi/2 - \Delta x/2$. This convention also applies to the other spatial directions $y^{\pm}$ and $z^{\pm}$.\\

To implement the symmetry boundary condition, we identify pairs of distribution function vectors that are symmetric counterparts. By applying appropriate transformations, such as reflections and rotations defined by transformation matrices, we map outgoing distribution functions at the boundaries to their symmetric incoming counterparts on the opposite side of the domain. This process ensures that the distribution functions interact correctly with the boundary nodes under the symmetry conditions. To account for symmetry in three dimensions, a set of transformation matrices is employed:

\begin{itemize}
    \item \textbf{Reflection Matrices $S_{\alpha}$}: Reverse the direction of the normal vector to the symmetry plane along the axis $\alpha$.
\end{itemize}
\renewcommand{\arraystretch}{0.9}
\begin{equation}
S_{x}=\begin{pmatrix}
-1 & 0 & 0\\
0 & 1 & 0\\
0 & 0 & 1
\end{pmatrix}, S_{y}=\begin{pmatrix}
1 & 0 & 0\\
0 & -1 & 0\\
0 & 0 & 1
\end{pmatrix}, S_{z}=\begin{pmatrix}
1 & 0 & 0\\
0 & 1 & 0\\
0 & 0 & -1
\end{pmatrix}
\end{equation}
\begin{itemize}
    \item \textbf{Rotation Matrices $R_{b}(\beta)$}: Rotate vectors around the axis $b$ by angle $\beta$.
\end{itemize}
\begin{equation}
R_{x}(\alpha)=\begin{pmatrix}
1 & 0 & 0\\
0 & \cos(\alpha) & -\sin(\alpha) \\
0 & \sin(\alpha)  & \cos(\alpha)
\end{pmatrix},
\end{equation}
\begin{equation}
R_{y}(\alpha)=\begin{pmatrix}
\cos(\alpha) & 0 & \sin(\alpha) \\
0 & 1 & 0\\
-\sin(\alpha)  & 0 & \cos(\alpha)
\end{pmatrix},
\end{equation}
\begin{equation}
R_{z}(\alpha)=\begin{pmatrix}
\cos(\alpha)  & -\sin(\alpha)  & 0\\
\sin(\alpha)  & \cos(\alpha)  & 0\\
0 & 0 & 1
\end{pmatrix}
\end{equation}
\begin{itemize}
    \item \textbf{Permutation Matrix $T_{xy}$}: Swap the $x$ and $y$ components of a vector\vspace{-.3cm}
\end{itemize}
\begin{equation}
T_{x,y}=\begin{pmatrix}
0 & 1 & 0\\
1 & 0 & 0\\
0 & 0 & 1
\end{pmatrix}
\end{equation}
The mapping of outgoing distribution function vectors to their symmetric counterparts involves identifying the indices of vectors that point outward from the symmetry planes, edges, or corners based on their direction components $d_i,\alpha$. Appropriate symmetry transformations such as reflections and rotations are applied using the transformation matrices as defined in the Tables \ref{tab:3D-Planes}, \ref{tab:3D-Edges} and \ref{tab:3D-Corners} to find the corresponding inward-pointing vectors on the shifted side of the domain. This process leverages the periodicity of the computational grid to shift the outgoing vectors to the positions of their symmetric counterparts without altering the underlying periodic computation of the algorithm.
\begin{table}[hb]
\centering
\caption{\textbf{Planes:} Displacement of the distribution function that match the direction on the side surfaces due to symmetry in three-dimensional space}
\label{tab:3D-Planes}
\begin{tabular}{lll}
\toprule
Displacement & Direction $c_i$ & Transf. \\
\midrule
$f_i(x, y^{+}, z) \to f_j(x, y^{-}, z)$ & $d_{i,y}=1$ & $S_y$ \\
$f_i(x, y, z^{+}) \to f_j(x, y, z^{-})$ & $d_{i,z}=1$ & $S_z$ \\
$f_i(x^{-}, y, z) \to f_j(x^{+}, y, z)$ & $d_{i,x}=-1$ & $S_x$ \\
$f_i(x^{+}, y, z) \to f_j(y, y^{-}, z)$ & $d_{i,x}=1$ & $R_z\left({3\pi}/{2}\right)$ \\
$f_i(x, y^{+}, z) \to f_j(x^{-}, x, z)$ & $d_{i,y}=1$ & $R_z\left({\pi}/{2}\right)$ \\
$f_i(x, y, z^{-}) \to f_j(y, x, z^{+})$ & $d_{i,z}=-1$ & $T_{xy},\ S_z$ \\
\bottomrule
\end{tabular}
\end{table}

\begin{table}[ht]
\centering
\caption{\textbf{Edges:} Displacement of the distribution functions due to symmetry at the edges in three-dimensional space}
\label{tab:3D-Edges}
\begin{tabular}{lll}
\toprule
Displacement & Vector direction $d_i$ & Transf. \\
\midrule
$f_i(x^{-}, y^{+}, z) \to f_j(x^{-}, y^{+}, z)$ & $d_{i,x} = -1$, $d_{i,y} = 1$ & $\mathbf{I}$ \\
$f_i(x^{+}, y^{-}, z) \to f_j(x^{+}, y^{-}, z)$ & $d_{i,x} = 1$, $d_{i,y} = -1$ & $\mathbf{I}$ \\
$f_i(x^{-}, y^{-}, z) \to f_j(x^{+}, y^{+}, z)$ & $d_{i,x} = -1$, $d_{i,y} = -1$ & $R_z(\pi)$ \\
$f_i(x^{+}, y^{+}, z) \to f_j(x^{-}, y^{-}, z)$ & $d_{i,x} = 1$, $d_{i,y} = 1$ & $R_z(\pi)$ \\
$f_i(x^{-}, y, z^{+}) \to f_j(x^{+}, x, z^{-})$ & $d_{i,x} = -1$, $d_{i,z} = 1$ & $R_y(\pi)$ \\
$f_i(x, y^{-}, z^{+}) \to f_j(x, y^{+}, z^{-})$ & $d_{i,y} = -1$, $d_{i,z} = 1$ & $R_x(\pi)$ \\
$f_i(x, y^{+}, z^{+}) \to f_j(x^{-}, x, z^{-})$ & $d_{i,y} = 1$, $d_{i,z} = 1$ & $R_z\left({\pi}/{2}\right),\ S_z$ \\
$f_i(x^{+}, y, z^{+}) \to f_j(y, y^{-}, z^{-})$ & $d_{i,x} = 1$, $d_{i,z} = 1$ & $R_z\left({3\pi}/{2}\right),\ S_z$ \\
$f_i(x, y^{+}, z^{-}) \to f_j(x, y^{-}, z^{+})$ & $d_{i,y} = 1$, $d_{i,z} = -1$ & $R_x(\pi)$ \\
$f_i(x, y^{-}, z^{-}) \to f_j(x^{+}, x, z^{+})$ & $d_{i,y} = -1$, $d_{i,z} = -1$ & $R_z\left({\pi}/{2}\right),\ S_z$ \\
$f_i(x^{-}, y, z^{-}) \to f_j(y, y^{+}, z^{+})$ & $d_{i,x} = -1$, $d_{i,z} = -1$ & $R_z\left({3\pi}/{2}\right),\ S_z$ \\
$f_i(x^{+}, y, z^{-}) \to f_j(x^{-}, y, z^{+})$ & $d_{i,x} = 1$, $d_{i,z} = -1$ & $R_y(\pi)$ \\
\bottomrule
\end{tabular}
\end{table}

\begin{table}[hbt]
\centering
\caption{\textbf{Corners:} Displacement of the distribution functions at the corners in three-dimensional space}
\label{tab:3D-Corners}
\begin{tabular}{lll}
\toprule
Displacement & Vector $d_i$ & Vector $d_j$ \\
\midrule
$f_i(x^{-}, y^{-}, z^{-}) \to f_j(x^{+}, y^{+}, z^{+})$ & $(-1, -1, -1)^\mathrm{T}$ & $(1, 1, 1)^\mathrm{T}$ \\
$f_i(x^{+}, y^{+}, z^{+}) \to f_j(x^{-}, y^{-}, z^{-})$ & $(1, 1, 1)^\mathrm{T}$ & $(-1, -1, -1)^\mathrm{T}$ \\
$f_i(x^{-}, y^{-}, z^{+}) \to f_j(x^{+}, y^{+}, z^{-})$ & $(-1, -1, 1)^\mathrm{T}$ & $(1, 1, -1)^\mathrm{T}$ \\
$f_i(x^{+}, y^{+}, z^{-}) \to f_j(x^{-}, y^{-}, z^{+})$ & $(1, 1, -1)^\mathrm{T}$ & $(-1, -1, 1)^\mathrm{T}$ \\
$f_i(x^{-}, y^{+}, z^{-}) \to f_j(x^{+}, y^{-}, z^{+})$ & $(-1, 1, -1)^\mathrm{T}$ & $(1, -1, 1)^\mathrm{T}$ \\
$f_i(x^{-}, y^{+}, z^{+}) \to f_j(x^{-}, y^{+}, z^{-})$ & $(-1, 1, 1)^\mathrm{T}$ & $(-1, 1, -1)^\mathrm{T}$ \\
$f_i(x^{+}, y^{-}, z^{+}) \to f_j(x^{+}, y^{-}, z^{-})$ & $(1, -1, 1)^\mathrm{T}$ & $(1, -1, -1)^\mathrm{T}$ \\
$f_i(x^{+}, y^{-}, z^{-}) \to f_j(x^{-}, y^{+}, z^{+})$ & $(1, -1, -1)^\mathrm{T}$ & $(-1, 1, 1)^\mathrm{T}$ \\
\bottomrule
\end{tabular}
\end{table}

In three dimensions, special attention must be given to vectors associated with edges and corners, as they involve combinations of two or all three direction components. Vectors pointing out along edges require transformations that account for the combination of direction components (e.g., $d_{i,x}=-1$$d_{i,y}=1$). Similarly, vectors pointing out from corners involve all three components and may necessitate combined transformations using reflections and rotations to accurately map them to their symmetric counterparts. Although some stencils, such as D3Q19, may not include vectors directly pointing out from corners, it is crucial to handle them appropriately if present to ensure accurate symmetric interactions.

\begin{figure}[hbt!]
    \centering
    \includegraphics[width=1.00\linewidth]{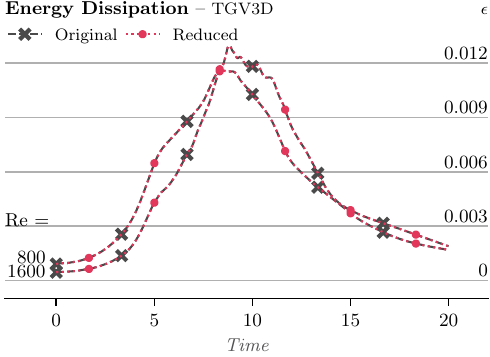}
    \caption{Comparison of the energy dissipation of the Taylor-Green vortex, demonstrating that equivalent results can be achieved using a reduced domain with symmetry boundary conditions.}
    \label{fig:tgv_appendinx}
\end{figure}

By systematically rearranging the distribution functions that point outward from the domain in each time step, we ensure that the symmetry boundary conditions are accurately represented. Figure \ref{fig:tgv_appendinx} compares the dissipation rates calculated using the \textsc{BGK} operator with the same discretization $\Delta x$. It can be seen that the results remain unchanged when simulating only a segment of the original domain, provided that the discussed boundary conditions are properly applied.

\bibliography{literature}
\bibliographystyle{unsrt}

\end{document}